\documentclass{SciPost}

\hypersetup{
    colorlinks,
    linkcolor={red!50!black},
    citecolor={blue!50!black},
    urlcolor={blue!80!black}
}

\usepackage[bitstream-charter]{mathdesign}
\DeclareSymbolFont{usualmathcal}{OMS}{cmsy}{m}{n}
\DeclareSymbolFontAlphabet{\mathcal}{usualmathcal}

\fancypagestyle{SPstyle}{
\fancyhf{}
\lhead{\colorbox{scipostblue}{\bf \color{white} ~SciPost Physics }}
\rhead{{\bf \color{scipostdeepblue} ~Submission }}

\fancyfoot[C]{\textbf{\thepage}}
}

\begin{document}
\pagestyle{SPstyle}
\nolinenumbers

\begin{center}{\Large \textbf{\color{scipostdeepblue}{
A multi-parameter expansion for the evolution of asymmetric binaries in astrophysical environments\\
}}}\end{center}

\begin{center}\textbf{
Sayak Datta\textsuperscript{1,2$\star$} and
Andrea Maselli\textsuperscript{1,2$\dagger$}
}\end{center}

\begin{center}
{\bf 1} Gran Sasso Science Institute (GSSI), I-67100 L’Aquila, Italy \\
{\bf 2} INFN, Laboratori Nazionali del Gran Sasso, I-67100 Assergi, Italy
\\[\baselineskip]
$\star$ \href{mailto:sayak.datta@gssi.it}{\small sayak.datta@gssi.it}\,,\quad
$\dagger$ \href{mailto:andrea.maselli@gssi.it}{\small andrea.maselli@gssi.it}
\end{center}

\section*{\color{scipostdeepblue}{Abstract}}
\textbf{\boldmath{
Compact binaries with large mass asymmetries - such as 
Extreme and Intermediate Mass Ratio Inspirals - are 
unique probes of the astrophysical environments in which 
they evolve. Their long-lived and intricate dynamics allow 
for precise inference of source properties, provided 
waveform models are accurate enough to capture the full 
complexity of their orbital evolution. In this work, we develop 
a multi-parameter formalism, inspired by vacuum perturbation 
theory, to model asymmetric binaries embedded in general 
matter distributions with both radial and tangential pressures. 
In the regime of small deviations from the Schwarzschild 
metric, relevant to most astrophysical scenarios, the system 
admits a simplified description, where both metric and fluid 
perturbations can be cast into wave equations closely related 
to those of the vacuum case. This framework offers a practical 
approach to modeling the dynamics and the  gravitational wave 
emission from binaries in realistic matter distributions, and can be modularly integrated with existing results for vacuum sources.
}}


\vspace{10pt}
\noindent\rule{\textwidth}{1pt}
\tableofcontents
\noindent\rule{\textwidth}{1pt}
\vspace{10pt}

Coalescing binaries with large mass asymmetry, i.e., mass 
ratios \( q \ll 1 \), represent a novel class of gravitational wave 
(GW) sources for next-generation detectors, as they remain 
undetectable by current interferometers. These systems consist 
of a stellar or an intermediate-mass compact object (the secondary) orbiting a significantly more massive black hole (the primary).

Among these, Extreme Mass Ratio Inspirals (EMRIs), where 
a primary of mass \( \sim (10^6 - 10^8) M_\odot \) pairs 
with a companion of \( \sim (10 - 10^2) M_\odot \), can 
be observed continuously for tens of thousands of orbits 
\cite{Berry:2019wgg}. During this phase, the secondary 
evolves within a few gravitational radii of the primary before 
the final plunge, emitting GWs that peak in the 
millihertz regime—well within LISA’s \cite{2017arXiv170200786A} or TianQin's \cite{TianQin:2015yph} sensitivity 
range.\footnote{Exotic scenarios, 
such as those involving sub-solar black holes, could allow 
EMRIs with primaries as light as \( 10^3 M_\odot \), making 
them potential targets for third-generation detectors, 
with GW emission frequencies below 10 Hz 
\cite{Barsanti:2021ydd,Miller:2020kmv}.}  
Intermediate Mass Black Holes (IMBHs), with masses in the 
range \( (10^2 - 10^4) M_\odot \), can form Intermediate Mass Ratio 
Inspirals (IMRIs) when coupled with either stellar-mass 
or supermassive black holes (BHs), with mass ratios \( q \sim 10^{-4} - 10^{-2} \) \cite{Klein:2015hvg, Volonteri:2020wkx}. 
IMRIs  have shorter inspirals and less variability than EMRIs \cite{Arca-Sedda:2020lso}, 
emitting GWs across a broad frequency range, from \( 10^{-3} \) Hz to 
10 Hz. This makes them multi-band sources, potentially detectable 
by mHz \cite{LISA:2024hlh,2025arXiv250220138L}, decihertz observatories \cite{Ajith:2024mie}, and 3G detectors \cite{Abac:2025saz,2023arXiv230613745E}.  

As  $q$  decreases, the inspiral duration and the number of GW cycles 
followed by asymmetric binaries increase significantly \cite{Barack:2018yvs}. 
These systems spend a substantial portion of their inspiral in a strong-field 
regime, tracing highly relativistic, eccentric, and off-equatorial trajectories 
before merging. The combination of such a large number of GW cycles 
and rich relativistic dynamics is crucial for achieving unprecedented 
precision in measuring source parameters \cite{Berry:2019wgg}, 
and advancing the fundamental physics science goals expected by 
GW observations of these systems \cite{Barausse:2020rsu,LISA:2022kgy,Sedda:2019uro,Cardenas-Avendano:2024mqp}. 

Asymmetric binaries have garnered increasing attention as 
prime sources for probing the astrophysical environments in 
which they evolve \cite{Barausse:2020rsu}. Indeed, BHs do not 
exist in isolation; they inhabit diverse environments where 
particles and fields, potentially of unknown or exotic nature, 
interact both with each other and with the compact objects. For 
instance, massive BHs are often surrounded by dark matter 
halos, which may consist of exotic fields or beyond-standard-model candidates \cite{Bertone:2018krk}. These surrounding 
structures can redistribute in the presence of a BH, forming overdensities that influence the binary’s orbital dynamics and imprint 
characteristic signatures on the emitted GW signals \cite{Barausse:2014tra,Cardoso:2019rou}. Such signals 
carry valuable information about changes in the galactic 
potential and local interactions, such as those arising 
from dynamical friction \cite{Eda:2013gg,Eda:2014kra,Kavanagh:2020cfn,Cardoso:2021wlq,Speeney:2022ryg,Gliorio:2025cbh,Kakehi:2025peb,Feng:2025fkc,Mitra:2025tag,Chandrasekhar:1943ys,Ostriker:1998fa}.

Moreover, crowded galactic centers can induce tidal resonances 
that influence EMRI evolution and reveal nearby stellar-mass 
object distributions \cite{Bonga:2019ycj}.  IMRIs are also expected 
to form in dense, matter-dominated environments, such as the accretion 
disks of active galactic nuclei \cite{Arca-Sedda:2020lso}. These 
systems interact with the surrounding gas through effects such as 
density wakes, gap-opening processes, and tidal torques, leading to 
complex GW emission patterns \cite{Garg:2022nko}. Observing such 
effects could constrain disk properties and enable multi-messenger 
analyses via electromagnetic counterparts \cite{Speri:2022upm}.  

Modeling GW emission from asymmetric binaries requires, however, 
highly accurate waveforms \cite{Babak:2006uv}. The self-force (SF) 
formalism provides the most precise framework to describe such systems, 
capturing their full evolutionary complexity \cite{Barack:2018yvs,Hinderer:2008dm}. 
In this approach, Einstein field equations are expanded in powers of the 
mass ratio \( q \). The leading-order solution models the secondary as a 
point particle moving along the geodesics of the primary, while higher-order 
corrections account for self-interaction and finite-size effects. On the 
radiation-reaction timescale, the GW phase evolution in the SF expansion follows:  
\begin{equation}
\varphi = \frac{\varphi^{(0)}}{q} + \varphi^{(1)} + q\varphi^{(2)} + \cdots,  
\label{eq:SFphaseevo}
\end{equation}  
where \( \varphi^{(0)} \) and \( \varphi^{(1)} \) correspond to the adiabatic (0PA) and 
post-adiabatic (1PA) contributions, respectively \cite{Hinderer:2008dm}. Phase accuracy 
at sub-radian levels is needed for precise parameter estimation, requiring calculations 
up to at least the 1PA order. The leading dissipative effects govern the 0PA phase evolution, 
while\footnote{Orbital resonances introduce additional corrections at the 0.5PA order \cite{Hinderer:2008dm}.} 
first-order conservative SF and second-order dissipative SF effects contribute to the 1PA phase 
component \( \varphi^{(1)} \). After nearly three decades of effort, recent work has achieved 
the first implementation of a 1PA waveform \cite{Pound:2019lzj,Wardell:2021fyy,Warburton:2021kwk}.  

Moving beyond vacuum General Relativity presents significant challenges due to the lack of 
relativistic solutions describing BHs embedded in matter and the complexities introduced 
by metric-matter couplings. As a result, modeling environmental effects on EMRIs often 
relies on post-Newtonian approaches \cite{Kocsis:2011dr,Kavanagh:2020cfn,Coogan:2021uqv,Cole:2022yzw,Tomaselli:2023ysb,Berezhiani:2023vlo,Macedo:2013qea}, 
though fully relativistic descriptions remain key to confidently extract small deviation 
from vacuum predictions 
\cite{Annulli:2020lyc,Traykova:2021dua,Destounis:2021mqv,Vicente:2022ivh,Speeney:2022ryg,Destounis:2022obl,Khalvati:2024tzz,Vicente:2025gsg,Mitra:2025tag,Yuan:2025fde,Santos:2025ass,Tomaselli:2024ojz}.  

Notable exceptions that provide ab initio background models incorporating non-vacuum 
contributions include studies investigating how ultra-light scalar fields surrounding massive 
primaries influence EMRI evolution at leading SF order \cite{Duque:2023seg,Brito:2023pyl}. 
A recent study built a relativistic perturbative framework for investigating EMRIs and IMRIs 
in dense environments, focusing on scalar clouds formed via superradiance around Kerr 
BHs \cite{Dyson:2025dlj,Li:2025ffh}, emphasizing the relevance of spin effects in assessing matter 
contributions to GW signals 

Along with fundamental physics motivations, scalar fields likely provide the most accessible 
framework for modeling environmental effects. Efforts to model the interaction of asymmetric 
binaries with generic fluids remain limited due to the complexity of the calculations. A fully 
relativistic approach, recently developed to model GW emission from EMRIs embedded 
in spherically symmetric matter distributions \cite{Cardoso:2022whc,Cardoso:2021wlq}, 
using both semi-analytical and fully numerical methods \cite{Figueiredo:2023gas,Rahman:2023sof,Speeney:2024mas,Gliorio:2025cbh, Rahman:2025mip}, 
revealed a rich and intricate phenomenology arising from a fully relativistic treatment. 
This model also underscored the significant increase in computational complexity due 
to matter components and their perturbations. As a result, even at 0PA, generating accurate 
waveforms across a broad parameter space remains unfeasible.

However, in most astrophysically relevant cases, and in the dynamical regimes of interest 
for GW detectors, environmental effects are expected to be “small”. In this regime, the background geometry of asymmetric binaries is dominated 
by the BH vacuum spacetime, in which both the companion and the surrounding matter 
act as perturbations, leading to substantial simplifications.

Following this path, we develop a multi-parameter framework 
to describe the evolution of asymmetric binaries embedded in 
generic, low-density environments, modeled via a fluid 
stress-energy tensor. We adopt a general anisotropic prescription that incorporates both radial and tangential pressure components. 
Focusing on non-spinning BHs, we solve Einstein equations by 
computing axial and polar perturbations at first order in the mass ratio. 
We provide practical, ready-to-use formulas for computing both gravitational and fluid perturbations, as well as the resulting 
GW emission at the adiabatic order, expressed in terms of 
environmental parameters and the secondary’s orbital trajectory.
Throughout this work, we use units in which $ G = c = 1$, unless speficied otherwise.

%
\section{Field equations and the Multi-parameter expansion}
%

Our starting point is the action for generic environmental 
fields $\vartheta$:
\begin{equation}\label{eq:action}
S=\int \frac{\sqrt{-g}}{16\pi}d^4x \, \mathcal{R}+ S_e[g_{\mu\nu},\vartheta]+S_p[g_{\mu\nu},\varphi]\ ,
\end{equation}
where the action $S_p$ describes the perturber secondary of mass 
$m_p$ and its internal matter fields $\varphi$, which can be treated 
using a skeletonized approach \cite{Damour:1992we}, 
$\mathcal{R}$ is the Ricci scalar, and $g$ the metric determinant.
The field equations for $g_{\mu\nu}$, can be derived by varying 
the total action with respect to the metric, that yields
\begin{equation}
G_{\mu\nu}=8\pi T^e_{\mu\nu}+
8\pi T^p_{\mu\nu}\ ,\label{eq:feildsG}
\end{equation}
where  $G_{\mu\nu}$ is the Einstein operator, and 
$T^{e,p}_{\mu\nu}$ are the stress-energy tensors related to the 
environment and the secondary, 
\begin{equation}
T_{\mu\nu}^{e,p}=-\frac{16\pi}{\sqrt{-g}}\frac{\delta \sqrt{-g}{\cal L}_{e,p}}{\delta g^{\mu\nu}}\ ,
\end{equation}
where ${\cal L}_{e,p}$ are the Lagrangian densities 
associated with the actions $S_{e,p}$. The total 
energy-momentum tensor satisfies the covariant equation 
\begin{equation}\label{eq:nablaT}
\nabla_{\mu} T^{\mu}{_{\nu}} = \nabla_{\mu} (T^{e\mu}{_{\nu}}+T^{p\mu}{_{\nu}})= 0\ . 
\end{equation}

We assume the primary is a BH of mass $M$ dressed by 
a stationary distribution of matter, with a stress-energy 
tensor for a generic anisotropic fluid\footnote{A 
prescription to describe anisotropic fluids in Newtonian 
gravity and in General Relativity has been recently proposed 
in \cite{Cadogan:2024ohj,Cadogan:2024mcl,Cadogan:2024ywc}, 
aiming to cure certain inconsistencies arising due to Eq.~\eqref{eq:envT} 
when modeling stellar solutions. Such formalism can in principle be adapted to our approach.}: 
\begin{equation}\label{eq:envT}
     T^{e}_{\mu\nu} = \rho u_{\mu}u_{\nu} + p_r k_{\mu}k_{\nu} + p_t \Pi_{\mu\nu}\ ,
\end{equation}
where we call $p_t$ and $p_r$ as radial and tangential pressures,
$u^\mu$ is the fluid four velocity and $k^\mu$ is a unit space-like 
radial vector orthogonal to the later, such that $-u_{\mu}u^{\mu} = k_{\mu}k^{\mu}=1$ 
and $u_{\mu}k^{\mu}=0$ \cite{Bowers:1974tgi,Doneva:2012rd,Raposo:2018rjn}. 
The projector on the surface orthogonal to the 4-velocity and $k^\mu$ 
is given by $\Pi_{\mu\nu}= g_{\mu\nu} + u_{\mu}u_{\nu} - k_{\mu}k_{\nu}$, 
with $u^\mu \Pi_{\mu\nu}X^\nu=k^\mu \Pi_{\mu\nu}X^\nu=0$, for a 
generic vector $X^\nu$. 

The secondary BH can be introduced with a perturbative approach, using 
the mass ratio $q = m_p / M \ll 1$ as parameter of the expansion. In this work we 
consider linear-order perturbations in $q$, which correspond to 
the leading dissipative contribution in a generic SF expansion 
of the binary dynamics \cite{Barack:2018yvs}. In this setup, the secondary 
evolves along a flow of geodesics driven by the energy and angular 
momentum fluxes. Higher-order terms, as well as a two-timescale analysis 
of environmental effects, will be studied elsewhere. 
The energy momentum of the secondary is given by:
\begin{equation}
    T^{p\mu\nu}(x^\alpha) = m_p \int_\gamma u_p^{\mu} u_p^{\nu} \frac{\delta^{(4)}(x^{\mu}-x_p^{\mu}(\tau))}{\sqrt{-g}}d\tau\ ,
\end{equation}
where $\gamma$ is the worldline of the compact object, $\tau$ its 
proper time, and $u_p^{\mu}(\tau) = dx_p^{\mu}/d\tau$ its $4-$ 
velocity.

We introduce a bookkeeping parameter $\epsilon$ to characterize 
the perturbative nature of the matter distribution, which will 
later guide the classification of environmental effects. With 
$\rho$ setting the scale of the environmental stress-energy 
tensor \eqref{eq:envT}, we follow \cite{Dyson:2025dlj} 
and define $\epsilon$ as the ratio between the environmental 
and BH densities, $\epsilon=(M_e/L_e^3)/(M/L^3)$, 
where $M_e$ and $L_e$ are the mass and the scale of the 
distribution, and $L\sim M$ the BH scale.
For instance, in the case of the dark matter configurations 
considered in \cite{Cardoso:2021wlq}, one finds $\epsilon = (M_{\rm halo}/M)/(a_0/L)^3$, with $M_{\rm halo}$ and $a_0$ denoting the halo mass and its typical size, respectively. 
In addition to density, the compactness of the matter distribution, defined as ${\cal C}_e=M_e/L_e$, is expected to play a central role in determining the behavior of perturbations \cite{Cardoso:2021wlq,Gliorio:2025cbh}. 
Expressing $\epsilon$ in terms of ${\cal C}_e$ one obtains $\epsilon\sim{\cal C}_e^3(M/M_e)^2$, suggesting that the perturbative treatment remains valid as long as ${\cal C}_e\lesssim(M_{e}/M)^{2/3}$. For example, for typical dark matter halos, with $M_e\sim (10^5-10^6)M$, the compactness satisfies ${\cal C}_e\ll 1$, ensuring 
$\epsilon\ll 1$.

When $\epsilon \sim \mathcal{O}(1)$, the background metric deviates significantly from the Kerr solution. Conversely, when 
$\epsilon \ll \mathcal{O}(1)$, environmental effects can be treated as small 
perturbations of the vacuum BH background, and the binary dynamics is 
governed by two small parameters: $\epsilon$ and the mass ratio $q$.

In this work, we focus on the latter regime and compute the equations 
describing metric and matter perturbations by expanding the field equations 
\eqref{eq:feildsG}, the covariant conservation of $T^\mu{_\nu}$ \eqref{eq:nablaT}, 
and all relevant tensor quantities in powers of $\epsilon$ and $q$. We retain terms 
up to $\mathcal{O}(\epsilon q)$, such that the metric and stress-energy tensors 
can be expressed as:
\begin{align}
    g_{\mu\nu} =& g_{\mu\nu}^{(0,0)} +q g_{\mu\nu}^{(1,0)} + \epsilon g_{\mu\nu}^{(0,1)} + q \epsilon g_{\mu\nu}^{(1,1)} \label{eq:metricexp},\\
    T^{e}_{\mu\nu} =& \epsilon T^{e}_{\mu\nu} {}^{(0,1)} + q \epsilon T^{e}_{\mu\nu} {}^{(1,1)}\quad\ ,\quad
    T^{p}_{\mu\nu} =  q T^{p}_{\mu\nu} {}^{(1,0)} + q \epsilon T^{p}_{\mu\nu} {}^{(1,1)}  ,
\end{align}
where superscripts $(i,j)$ identify the expansion order 
$\mathcal{O}(q^i, \epsilon^j)$.
In the limit $\epsilon\rightarrow0$ the formalism 
reduces to a particle moving in the Schwarzschild spacetime, with 
perturbations described by the Regge-Wheeler-Zerilli equations 
\cite{Regge:1957td,Zerilli:1970wzz,Zerilli:1970se}.

To isolate the various contributions at orders $\epsilon$ 
and $q$, we expand the nonlinear Einstein tensor 
$G_{\mu\nu}[g_{\alpha\beta}]$ about the background $g^{(0,0)}_{\alpha\beta}$, as in Eq.~\eqref{eq:metricexp}. 
For a generic perturbation $h_{\alpha\beta}$, we define 
the $n$-th variations by
\begin{equation}
G^{[n]}_{\mu\nu}[h_{\alpha\beta}] = \frac{1}{n!}\left.\frac{d^n}{d\lambda^n}\,
G_{\mu\nu}\!\big[g^{(0,0)}_{\alpha\beta}+\lambda\,h_{\alpha\beta}\,\big]\right|_{\lambda=0}.
\end{equation}
Then
\begin{equation}
G_{\mu\nu}\![g^{(0,0)}_{\alpha\beta}+h_{\alpha\beta}]
=
G_{\mu\nu}\![g^{(0,0)}]
+ G^{[1]}_{\mu\nu}[h_{\alpha\beta}]
+ G^{[2]}_{\mu\nu}[h_{\alpha\beta},h_{\alpha\beta}]
+ G^{[3]}_{\mu\nu}[h_{\alpha\beta},h_{\alpha\beta},h_{\alpha\beta}]
+\dots \, .
\label{eq:Gexp}
\end{equation}
Inserting the metric expansion \eqref{eq:metricexp} 
into Eq.~\eqref{eq:Gexp} and keeping terms up to mixed 
order $\mathcal{O}(\epsilon q)$ yields
\begin{align}
G_{\mu\nu}[g_{\alpha\beta}]
&=
G_{\mu\nu}\![g^{(0,0)}]
+ \epsilon\, G^{[1]}_{\mu\nu}\!\big[g^{(0,1)}\big]
+ q\, G^{[1]}_{\mu\nu}\!\big[g^{(1,0)}\big]
\notag\\
&\quad
+ \epsilon q \Big(
G^{[1]}_{\mu\nu}\!\big[g^{(1,1)}\big]
+ G^{[2]}_{\mu\nu}\!\big[g^{(1,0)},\,g^{(0,1)}\big]
\Big)
\label{eq:Gexp2}
\end{align}
We assume the background solves the zeroth-order 
field equations,
$G_{\mu\nu}[g^{(0,0)}]=0$, which 
in Schwarzschild coordinates
$x^\mu=(t,r,\theta,\phi)$ gives
\begin{equation}
g^{(0,0)}_{\mu\nu}=\mathrm{diag}\!\left(-f,\,f^{-1},\,r^2,\,r^2\sin^2\theta\right),
\qquad f=1-\frac{2M}{r}\, .
\end{equation}
For clarity, in what follows we absorb the explicit factors of $q$ and $\epsilon$ within each term 
of the expansion.\\

The perturbative framework developed above is valid
when both the amplitude of the environmental effects
and the contribution of the secondary remain small,
i.e.\ for $\epsilon \ll 1$ and $q \ll 1$.
Within this regime, nonlinear backreaction on the
background geometry is perturbative, 
and all quantities in Eqs.~\eqref{eq:metricexp}--\eqref{eq:Gexp2} can be consistently
expanded in powers of these parameters.

Moreover, we can estimate the regime in which nonlinear
hydrodynamic effects within the fluid may become relevant
by introducing an additional, although Newtonian, physical
scale that controls the strength of the local fluid
response. In our spherically symmetric configuration, the
Bondi--Hoyle--Lyttleton radius $r_B$ \cite{Bondi:1952ni}
provides a diagnostic of the region where the surrounding
fluid becomes gravitationally bound to the secondary and
nonlinear effects may arise.
For orbits at radius $r = x\,M$, with $x$ 
the dimensionless orbital separation in units 
of the primary mass $M$, the ratio 
$r_B/r \sim q/[x(c_s^2 + v_{\rm rel}^2)]$ 
remains well below unity for typical EMRIs 
($q \sim 10^{-5}$) whenever either the sound speed 
$c_s$ or the relative velocity $v_{\rm rel}$ 
between the fluid and the secondary exceeds a few 
$10^{-3}c$, where $c$ is the speed of light \cite{Goodman:2000zg,Cimerman:2021,Ono:2025}.
This condition is naturally met in warm or hot subsonic flows, ensuring that the fluid response 
stays in the linear regime and that the 
point-particle approximation holds.

%
\section{Solutions of the multi-parameter expansion}

%
\subsection{Environmental effects: (0,1) contributions}
%

The \((0,1)\) corrections to the metric tensor satisfy the 
inhomogeneous equations  
\begin{eqnarray}
G^{[1]\mu}{_{\nu}}\!\big[g^{(0,1)}\big] = 8\pi T^{e\mu}{_\nu}{}^{(0,1)}\ .
\end{eqnarray}
To determine the components of the environmental 
stress-energy tensor, we utilize the normalization and orthogonality properties of the fluid four-velocity 
and the vector \( k^\mu \). For a stationary fluid 
with $u^\mu = (u^t,0,0,0)$ and $k^\mu = (k^t, k^r, 0, 0)$, 
these conditions lead to  
\begin{equation}\label{eq:eq01}
u^{t} = (-g_{tt})^{-1/2}, \quad k^t = 0, \quad k^r = g_{rr}^{-1/2}\ .
\end{equation}
Expanding the metric and matter variables in powers 
of \(\epsilon\), we obtain the explicit form of 
\( T^{e\mu}{_\nu}{}^{(0,1)} \):  
\begin{align}
\label{eq:environment-EMT01}
T^{e\mu}{_\nu}{}^{(0,1)} = \text{diag}(-\rho^{(0,1)}, p_r^{(0,1)}, p_t^{(0,1)}, p_t^{(0,1)})\ .
\end{align}
For sake of clarity, hereafter we drop the suffix 
$(0,1)$ from the background pressure and density 
functions.

At order \((0,1)\), we assume the following ansatz for the metric components:
\begin{equation}
g_{\mu\nu}^{(0,1)} = \text{diag}\left(-f H, \frac{2m}{r f^{2}}, 0,0\right)\ ,
\end{equation}
where both $H(r)$ and $m(r)$ are functions of the radial coordinate $r$ 
only. We focus on asymptotically flat solutions for which the matter 
variables vanish at the BH horizon $r_h$. This condition fixes $r_h = 2M$, 
as in the vacuum case, given that $m(r_h) = 0$.
At spatial infinity, the functions behave as
$H(r \rightarrow \infty) = -2M_e/r + \mathcal{O}\left(1/r^2\right)$
and $m(r \rightarrow \infty) = M_e + \mathcal{O}\left(1/r\right)$, 
such that
$g_{tt}(r \rightarrow \infty) = -1 + 2(M + M_e)/r\ ,$
where \(M + M_e\) is the total ADM mass of the system, and \(M_e\) denotes the mass of the matter distribution.

From the \(tt\) and \(rr\) components of Eq.~\eqref{eq:eq01}, 
we derive two ordinary differential equations  for $H$ and $m$:
\begin{equation}\label{eq:01Hm}
\frac{d m}{dr} = 4\pi r^2 \rho\quad \ ,\quad 
\frac{r^2 f^2}{2} \frac{dH}{dr} = m + 4\pi r^3 f p_r\ .
\end{equation}
Additionally, the energy-momentum covariant derivative 
at order \((0,1)\) gives:
\begin{equation}\label{eq:01pr}
\frac{dp_r}{dr} = \frac{2 }{r}p_t + \frac{(3 M - 2 r) }{r^2 f}p_r - \frac{M }{r^2 f}\rho\ .
\end{equation}
Equations~\eqref{eq:01Hm}-\eqref{eq:01pr} alone do not fully determine 
a solution for the metric and fluid variables. For a given density profile 
$\rho(r)$, which depends on the specific matter distribution, 
additional equations are required to close the system. This is typically 
provided by an equation of state that relates $p_r, p_t$, 
and $\rho$, and that we assume to be barotropic.  

The background metric $g_{\mu\nu}^{(0,0)}+g_{\mu\nu}^{(0,1)}$ allows for 
the study of the geodesic properties of both massless and massive particles. 
For example, the energy and angular momentum per unit mass, 
$({\cal E},{\cal L})$, of a massive body on a circular orbit of radius $r_p$ are given by:
\begin{align}
{\cal E}&={\cal E}^{(0,0)}+
\frac{f_p[(1-4f_p+3f_p^2)H_p-2f_pM H'_p]}{\sqrt{2}(f_p-1)(3f_p-1)^{3/2}}\ ,\label{eq:circenergy}\\
{\cal L}&={\cal L}^{(0,0)}
+\frac{4f_p^2M^2H'_p}{(1-f_p)^{5/2}(3f_p-1)^{3/2}}\ .\label{eq:circang}
\end{align}
where the vacuum expressions read:
\begin{equation}
{\cal E}^{(0,0)}=\frac{\sqrt{2}f_p}{(3f_p-1)^{1/2}}\ \ \ , \ 
{\cal L}^{(0,0)}=\frac{2M}{(4f_p-3f_p^2-1)^{1/2}}\ ,
\end{equation}
and $f_p=1-2M/r_p$, $H_p=H(r_p)$, $H'_p=H'(r)\vert_{r=r_p}$
The corresponding angular frequency of the body up to the linear 
order in $\epsilon$ is:
\begin{equation}
\Omega_p=\frac{M^{1/2}}{r_p^{3/2}}+\frac{2MH_p+r_p(r_p-2M)H'_p}{4\sqrt{M}r_p^{3/2}}\ .
\end{equation}
%
%
\subsection{The motion of the secondary: (1,0)+(1,1) contributions}\label{sec:10equations}
%

The motion of the secondary generates time dependent 
perturbations on both the metric and the matter fields, 
at the linear order in the mass ratio. For technical 
reasons, that will be clear at the 
end of this section, we will treat the left-hand side of Einstein equations 
working with a single background perturbation tensor
\begin{equation}
    \delta g_{\mu\nu} = g_{\mu\nu}^{(1,0)} + g_{\mu\nu}^{(1,1)}\ , 
\end{equation}
which solves the linearised field's 
equations:
\begin{equation}
    G^{\mu}{_\nu}[\delta g_{\alpha\beta}] = 8\pi(T^{p\mu}{_\nu}{}^{(1,0)} + T^{e\mu}{_\nu}{}^{(1,1)}+T^{p\mu}{_\nu}{}^{(1,1)})\ .\label{eq:fields10}
\end{equation}

Since the decoupling of the vacuum $(1,0)$ and matter $(1,1)$ sectors is performed at the end of the procedure, the operator on the left-hand side of Eq.~\eqref{eq:fields10} implicitly includes the terms appearing in the expansion~\eqref{eq:Gexp2}, namely the linear operators $G^{[1]\mu}{_\nu}[g^{(1,0)}]$ and $G^{[1]\mu}{_\nu}[g^{(1,1)}]$, together with the mixed second-order contribution $G^{[2]\mu}{_\nu}[g^{(1,0)}_{\alpha\beta},g^{(0,1)}_{\alpha\beta}]$.

Given the symmetry of the background, metric perturbations 
can be separated into the usual families of axial ($A$) and polar ($P$) 
components \cite{Regge:1957td,Zerilli:1970wzz,Zerilli:1970se}:
\begin{equation}
\delta g_{\mu\nu}(x^\alpha) = \delta g_{\mu\nu}^{A}(x^\alpha) +\delta  g_{\mu\nu}^{P}(x^\alpha)\ .
\end{equation}
Axial and polar modes change sign as $(-1)^{\ell+1}$ and $(-1)^{\ell}$ 
under the coordinate inversion $(\theta\rightarrow \pi-\theta,\phi\rightarrow \phi+\pi)$, 
respectively. The two classes of perturbations decouple, and can be treated independently. 
We can expand $g_{\mu\nu}^{A}(x^\alpha)$ and $g_{\mu\nu}^{P}(x^\alpha)$ 
in a complete set of tensor harmonics, such that:
%
\begin{align}
\delta g_{\mu\nu}^{A} =&  \sum_{\ell,m} \frac{\sqrt{2\lambda}}{r} \left[ ih_{1,\ell m}(t,r) {\bf c}_{\ell m}(\theta,\phi) -h_{0,\ell m}(t,r) {\bf c}^0{}_{\ell m}(\theta,\phi)   
+\frac{\sqrt{\Lambda}}{r}h_{2,\ell m}(t,r){\bf d}_{\ell m}(\theta,\phi)
    \right]\ ,\label{eq:axial10}\\
\delta g_{\mu\nu}^{P} =& \sum_{\ell,m}\Big[-g_{tt} H_{0,\ell m} (t,r) {\bf a}^0{}_{\ell m}(\theta,\phi) - i \sqrt{2} H_{1,\ell m} (t,r) {\bf a}^1{}_{\ell m}(\theta,\phi)
- \frac{i}{r}\sqrt{2\lambda} \eta_{0,\ell m} (t,r) {\bf b}^0{}_{\ell m}(\theta,\phi)\nonumber\\
&+\frac{\sqrt{2\lambda}}{r} \eta_{1,\ell m} (t,r) {\bf b}{}_{\ell m}(\theta,\phi)+g_{rr}H_{2,\ell m} (t,r) {\bf a}{}_{\ell m}(\theta,\phi)
+\sqrt{\Lambda \lambda}G_{\ell m} (t,r)  {\bf f}_{\ell m}(\theta,\phi)\nonumber\\
&+ \left(\sqrt{2}  K_{\ell m} (t,r)-\frac{\lambda}{\sqrt{2}}G_{\ell m} (t,r)\right) {\bf g}_{\ell m}(\theta,\phi)
\Big]\ ,\label{eq:polar10}
\end{align}
%
where $\lambda=\ell(\ell+1)$, $\Lambda=(\ell+2)(\ell-1)/2$, and 
the sum over the multipolar indices $(\ell,m)$ runs from $\ell=0,\ldots,\infty$ and 
$m=-\ell,\ldots,\ell$. The ten basis components $\{{\bf c}^{\ell m}_{\mu\nu},{\bf c}^{0\ell m}_{\mu\nu}\ldots {\bf g}^{\ell m}_{\mu\nu}\}$ 
depend on the spherical harmonics $Y_{\ell m}(\theta,\phi)$ and their derivatives 
(see e.g. Appendix~A of \cite{Sago:2002fe} for their explicit expression). 
Among the ten unknown functions $\{h_{1 \ell m} \ldots K_{\ell m}\}$, the 
axial term $h_{2\ell m}$ and the three polar components 
$\{\eta_{0\ell m},\eta_{1\ell m},G_{\ell m}\}$ 
can be set to zero by adopting the Regge-Wheeler-Zerilli gauge, 
such that the metric satisfy
\begin{align}
\label{eq:RWZ-gauge}
\delta g_{\theta\phi}=0\ \ , \ \delta g_{\phi\phi}=\delta g_{\theta\theta}\sin^2\theta\ ,\nonumber\\
\partial_{\phi}(\delta g_{t \phi}/\sin\theta)+\partial_{\theta}(\delta g_{t \theta}/\sin\theta)=0\ ,\nonumber\\
\partial_{\phi}(\delta g_{r \phi}/\sin\theta)+\partial_{\theta}(\delta g_{r \theta}/\sin\theta)=0\ .
\end{align}

Similarly to the metric perturbations, we decompose the particle stress-energy 
tensor in the basis of tensor harmonics: 
\begin{align}
\label{eq:PP-EMT10}
    T^{p}_{\mu\nu}{}^{(1,0)} =\sum_{\ell,m}\Big[&\mathcal{A}^{0(1,0)}_{\ell m} {\bf a}^0{}_{\ell m}(\theta,\phi) +\mathcal{A}^{1(1,0)}_{\ell m} {\bf a}^1{}_{\ell m}(\theta,\phi) + \mathcal{A}^{(1,0)}_{\ell m} {\bf a}{}_{\ell m}(\theta,\phi) + \mathcal{B}^{0(1,0)}_{\ell m}{\bf b}^0{}_{\ell m}(\theta,\phi) \nonumber\\
    &+ \mathcal{B}^{(1,0)}_{\ell m}{\bf b}{}_{\ell m}(\theta,\phi)+\mathcal{Q}^{(1,0)}_{\ell m} {\bf c}_{\ell m}(\theta,\phi)+\mathcal{Q}^{0(1,0)}_{\ell m} {\bf c}^0{}_{\ell m}(\theta,\phi) +  \mathcal{D}^{(1,0)}_{\ell m} {\bf d}_{\ell m}(\theta,\phi)\nonumber\\
    &+  \mathcal{G}^{(1,0)}_{\ell m} {\bf g}_{\ell m}(\theta,\phi) +  \mathcal{F}^{(1,0)}_{\ell m} {\bf f}_{\ell m}(\theta,\phi) \Big]\ .
\end{align}
The specific form of the coefficients $\{\mathcal{A}^{0(1,0)}_{\ell m},\ldots \mathcal{F}^{(1,0)}_{\ell m}\}$ depends on the 
secondary orbital configurations. Finally, the form 
$T^p_{\mu\nu}{}^{(1,1)}$ can be constructed using the 
same ansatz of  Eqs.~\eqref{eq:PP-EMT10}, and replacing 
the functions with the correct order of the expansion, 
e.g. $\mathcal{Q}^{(1,0)}_{\ell m}\rightarrow \mathcal{Q}^{(1,1)}_{\ell m}$ (see Appendix~\ref{app:stressenergy} for further 
details).

%
\subsubsection{Environmental effects in the presence of 
the secondary: (1,1) matter decompositions}
%

The last piece of the multi-parameter expansion is given 
by the $(1,1)$ perturbations of matter energy-momentum tensor, 
$T^{e\mu}{_\nu}{}^{(1,1)}$. The covariant equations~\eqref{eq:nablaT} 
are determined, at this order, by three contributions:
\begin{equation}
\label{eq:EMT-Conservation11}
\nabla_{\mu}[g_{\alpha\beta}^{(0,0)}] (T^{e\mu}{_{\nu}}{}^{(1,1)}+T^{p\mu}{_{\nu}}{}^{(1,1)})+
\nabla_{\mu}[g_{\alpha\beta}^{(1,0)}] T^{e\mu}{_{\nu}}{}^{(0,1)}+\nabla_{\mu}[g_{\alpha\beta}^{(0,1)}]T^{p\mu}{_{\nu}}^{(1,0)}= 0\ \ , 
\end{equation}
where we identify with $\nabla_{\mu}[g_{\alpha\beta}^{(m,n)}]$ 
the covariant derivative depending on the metric at the $(i,j)$ 
order. The $(1,1)$ contributions to the matter stress-energy 
tensor depend on the energy density and the pressure perturbations:
\begin{eqnarray}
    \rho =& \rho(r) + \rho^{(1,1)}(t,r,\theta,\phi)\ ,\\
    p=& p_r(r) + p_r^{(1,1)} (t,r,\theta,\phi)\ ,\\
    p_t =& p_t(r) + p_t^{(1,1)} (t,r,\theta,\phi)\ .
\end{eqnarray}

We exploit again the symmetry of the background to separate 
angular and time-radial variables. We expand fluid variables 
in terms of standard spherical harmonics
\begin{align}
    \rho^{(1,1)}=& \sum_{\ell, m} \rho^{(1,1)}_{\ell m}(t,r) Y_{\ell m}(\theta,\phi)\ , \\
    p_r^{(1,1)} =&   \sum_{\ell, m} p^{(1,1)}_{r,\ell m}(t,r) Y_{\ell m}(\theta,\phi)
     ,\\
    p_t^{(1,1)}=&  \sum_{\ell, m} p^{(1,1)}_{t,\ell m}(t,r) Y_{\ell m}(\theta,\phi)\ .
\end{align}
Moreover, pressure and density perturbations are 
linked by an equation of state\footnote{Note 
that, since the physical properties of matter are not 
altered by linear perturbations, the underlying 
equation of state is assumed to remain unchanged.}, 
such that:
\begin{equation}
p^{(1,1)}_{r,\ell m} = c^2_{r,{\ell m}}(r) \rho^{(1,1)}_{\ell m}\quad \ , \quad p^{(1,1)}_{t,\ell m} = c^2_{t,{\ell m}}(r) \rho^{(1,1)}_{\ell m}\ ,\label{eq:csdef}
\end{equation}
where the tangential $(c^2_{t,{\ell m}})$ and the radial $(c^2_{r,{\ell m}})$ 
sound speeds are in general not constant, and are 
functions of the radial coordinate (See Ref.~\cite{Datta:2023zmd} for specific examples). 

Perturbations of the fluid velocity $u^\mu$ and $k^\mu$ 
can be written in terms of vector harmonics \cite{PhysRevD.46.4289}. 
Given the form of the matter stress-energy tensor in 
Eq.~\eqref{eq:envT} and  that, to leading order, the 
energy and pressure variables are $\mathcal{O}(\epsilon)$,
we only need terms of the order 
$u^{\mu(1,0)}$ and $k^{\mu(1,0)}$ to determine 
$T^{e\mu}{_{\nu}}{}^{(1,1)}$. 
The normalization of the 4-velocity reduces the independent 
component of the perturbations to three unknown functions. 
The explicit form of $u^{\mu(1,0)}$ and $k^{\mu(1,0)}$ is given by:
\begin{align}
    u^{t(1,0)}&= \frac{1}{2\sqrt{f}} \sum_{\ell,m}H^{(1,0)}_{0,\ell m}(t,r) Y_{\ell m}(\theta,\phi)\ , \\
    u^{r(1,0)} &= \frac{f^{3/2}}{4\pi  }\sum_{\ell,m}W^{(1,0)}_{\ell m}(t,r) Y_{\ell m}(\theta,\phi)\ , \\
    u^{\theta(1,0)} &= \frac{\sqrt{f}}{4\pi  r^2}\sum_{\ell,m}\left[V^{(1,0)}_{\ell m}(t,r) \partial_{\theta}- \frac{U^{(1,0)}_{\ell m}(t,r)}{\sin \theta} \partial_{\phi} \right]Y_{\ell m}(\theta,\phi)\ ,\\
    u^{\phi(1,0)} &= \frac{\sqrt{f}}{4\pi  r^2\sin^2\theta}\sum_{\ell,m}\left[V^{(1,0)}_{\ell m}(t,r) \partial_{\phi}  + U^{(1,0)}_{\ell m}(t,r)\sin \theta\partial_{\theta} \right]Y_{\ell m}(\theta,\phi)\ .
\end{align}
The form of $k^{r(1,0)}$ and $k^{t(1,0)}$ can be found 
using nomalisation and orthogonality conditions.

%
\section{Perturbation equations}
%

The procedure for determining axial and polar 
perturbations closely follows the vacuum case, 
which has been extensively studied in the literature 
since the seminal works of Regge and Wheeler 
\cite{Regge:1957td,Zerilli:1970wzz} and Zerilli 
\cite{Zerilli:1970se}. In this section, we revisit 
the key steps for deriving the master equations 
governing the evolution of $\delta g_{\mu\nu}^{(A,P)}$, 
and for isolating the contributions arising from 
the $(1,0)$ and $(1,1)$ terms. We refer the reader 
to Appendix~\ref{app:scaling} for further details 
on our initial assumption of working with a single 
metric perturbation in $q$, and on the decoupling 
between vacuum and matter components. We present 
most of the equations in a compact form, emphasizing 
their functional dependence on the metric and fluid 
perturbations. The full explicit expressions are 
provided in the accompanying \texttt{Mathematica} 
supplementary file \cite{SGREP_REPO}.

%
\subsection{$\ell \ge 2$ axial modes}
%

In the axial case, we use the $\theta\theta$ and $\phi\phi$ 
components of Eqs.~\eqref{eq:fields10} to express
the time derivative $\partial_t h_{0\ell m}$ as a function 
of $h_{1,\ell m}$ and $\partial_r h_{1,\ell m}$. 
Substituting the latter into the $r\theta$ component of 
Einstein equations and introducing the master variable 
$\bar{\phi}_{\ell m}=-h_{1,\ell m}/r(-g_{tt}/g_{rr})^{1/2}$, 
we obtain a single, second-order partial differential 
equation of the form:
\begin{align}
(-g_{tt}/g_{rr})\partial^2_{r}\bar{\phi}_{\ell m}-\partial^2_{t}\bar{\phi}_{\ell m}
+ a_1\partial_{r}\bar{\phi}_{\ell m} + a_2 \bar{\phi}_{\ell m} = S_{\ell m}\ , 
\label{eq:masterAx1}
\end{align}
where $a_{1,2}$ depend only on $(0,0)$ and $(0,1)$ 
quantities. The source term $S_{\ell m}$ contains 
contributions from the particle’s stress-energy tensor 
and the background fluid variables. We now introduce a 
new master function:
\begin{equation}
\phi_{\ell m}(t,r) = \sqrt{Z(r)}\bar{\phi}_{\ell m}(t,r)\ , \label{eq:Zmap}
\end{equation}
where $Z = f^{-1}(-g_{tt}/g_{rr})^{1/2}$. For 
environmental effects that 
can be treated as small perturbations of the Schwarzschild 
metric, as considered here, we write $Z(r) = 1 + \delta Z(r)$, 
where $\delta Z(r)$ is of order $\mathcal{O}(\epsilon)$:
\begin{equation}
\delta Z(r) = \frac{H(r)}{2} - \frac{m(r)}{r f}\ .\label{eq:Zmap}
\end{equation}
In terms of the new field $\phi_{\ell m}$, Eq.~\eqref{eq:masterAx1} 
becomes:
\begin{align}
f \partial_r(f \partial_r \phi_{\ell m}) + (1 - \delta Z) 
\partial^2_{t} \phi_{\ell m}+ a_3 \phi_{\ell m} = \bar{S}_{\ell m}\ , \label{eq:masterAx2}
\end{align}
At this point we can decompose the perturbation into vacuum 
and matter components, 
i.e., $\phi_{\ell m} = \phi^{(0,0)}_{\ell m} + \phi^{(1,1)}_{\ell m}$. 
Furthermore, by introducing the usual tortoise coordinate 
$r_\star$, such that $\partial_{r_*} = f\partial_r$,
we can eliminate the first radial derivative of the metric perturbations, 
obtaining the following wave equations:
\begin{align}
[\partial^2_{r_\star} - \partial^2_{t} - V^{ A}]\phi^{(1,0)}_{\ell m}(t,r) &= S^{A(1,0)}_{\ell m}(t,r)\ , \label{eq:ax01} \\
[\partial^2_{r_\star} - \partial^2_{t} - V^{ A}]\phi^{(1,1)}_{\ell m}(t,r) &= S^{A(1,1)}_{\ell m}(t,r)\ . \label{eq:ax11}
\end{align}
Thus, at linear order in $\mathcal{O}(\epsilon)$, the 
axial perturbation problem reduces to solving two wave 
equations, with the same scattering potential, which 
matches the Regge–Wheeler expression for the vacuum 
case:
\begin{equation}
V^{\rm A} = f \left( \frac{\ell(\ell+1)}{r^2} - \frac{6M}{r^3} \right)\ .
\end{equation}
The source $S^{A(1,0)}_{\ell m}(t,r)$ only depends 
on the coefficients of $T^{p(1,0)}_{\mu\nu}$ in 
Eq.~\eqref{eq:PP-EMT10}. The source $S^{A(1,1)}_{\ell m}$ 
is proportional to $T^{p(1,1)}_{\mu\nu}$, and contains 
contributions from the vacuum master function 
$\phi^{(1,0)}_{\ell m}$, multiplied by the matter density 
and pressure. Once Eqs.~\eqref{eq:ax01}–\eqref{eq:ax11} 
are solved, we can use Eq.~\eqref{eq:Zmap} to obtain 
the $(1,0)$ and $(1,1)$ components of $\phi_{\ell m}$, 
and consequently the expansion for the metric 
functions $h_{1,\ell m}=h^{(1,0)}_{1,\ell m}+h^{(1,1)}_{1,\ell m}$ 
and $h_{0,\ell m}=h^{(1,0)}_{0,\ell m}+h^{(1,1)}_{0,\ell m}$. 
Their explicit expressions are given in 
Appendix~\ref{app:metricrecon}. 

Finally, the velocity perturbation $U^{(1,0)}_{\ell m}$ 
can be derived from the $t\theta$ component of Einstein 
equations, which yields an algebraic relation between 
this quantity and the metric variables:
\begin{align}
\partial_t U^{(1,0)}_{\ell m}(t,r)=& -\frac{4\pi \partial_t h^{(1,0)}_{0,\ell m}}{f} + \frac{4\pi (3-2r)(\kappa_r-\kappa_t)h^{(1,0)}_{1,\ell m}}{r^2 \kappa_t} +S_{\ell m}^{U}(t,r)\ ,
\end{align}
with $S_{\ell m}^{U}(t,r)$ depending on the point particle motion, \(\kappa_t=\rho(r)+p_t(r)\), and \(\kappa_r=\rho(r)+p_r(r)\). Axial perturbations do not couple, at this order, to the energy density or pressure perturbations because of parity considerations.

%
\subsubsection{The frequency domain solution}
%

In the frequency domain, Eqs.~\eqref{eq:ax01}–\eqref{eq:ax11} 
reduce to two ordinary differential equations in the radial 
coordinate:
\begin{align}
[\partial^2_{r_\star} + \omega^2 - V^{ A}]\phi^{(1,0)}_{\ell m}(\omega,r) &= S^{A(1,0)}_{\ell m}(\omega,r)\ , \label{eq:ax01F} \\
[\partial^2_{r_\star} + \omega^2 - V^{ A}]\phi^{(1,1)}_{\ell m}(\omega,r) &= S^{A(1,1)}_{\ell m}(\omega,r)\ , \label{eq:ax11F}\end{align}
where, for a generic function $X(t,r)$:
\begin{equation}
X(\omega,r) =\frac{1}{2\pi} \int_{-\infty}^{+\infty} e^{i\omega t} X(t,r)\, dt\quad ,\quad 
X(t,r) = \int_{-\infty}^{+\infty} e^{-i\omega t} X(\omega,r)\, d\omega\ .
\end{equation}
Equations~\eqref{eq:ax01F}–\eqref{eq:ax11F} can be solved using 
a Green's function approach. We first solve the associated homogeneous 
equations with purely ingoing (-) boundary conditions at the horizon and 
purely outgoing (+) conditions at infinity:
\begin{align}
\phi^{(1,0)(-)}_{\ell m} \sim
\begin{cases}
 e^{-i\omega r_\star} \quad &  r_\star \to -\infty\  \\
 A_{in}e^{-i\omega r_\star}+A_{\rm out}e^{i\omega r_\star} \quad &  r_\star \to +\infty \\
\end{cases}\ ,
\end{align}
\begin{align}
\phi^{(1,0)(+)}_{\ell m} \sim
\begin{cases}
 e^{i\omega r_\star} \quad &  r_\star \to +\infty\  \\
 B_{\rm in}e^{-i\omega r_\star}+B_{\rm out}e^{i\omega r_\star} \quad &  r_\star \to -\infty \\
\end{cases}\ ,
\end{align}
Note that the homogeneous equation is identical for both the 
$(1,0)$ and $(1,1)$ components, and hence needs to be solved 
only once. The full solution is obtained by integrating 
$\phi^{(1,0)(\pm)}_{\ell m}$ over the source term:
\begin{equation}
\phi^{(1,0)}_{\ell m} = {C}^{+} \phi^{(1,0)(-)}_{\ell m} + {C}^{-} \phi^{(1,0)(+)}_{\ell m}\ , \label{eq:axialfullF}
\end{equation}
with coefficients given by:
\begin{equation}
{C}^{+} = \int_{-\infty}^{r_\star} \frac{\phi^{(1,0)(-)}_{\ell m}(r'_\star)\, S^{(1,0)}_{\ell m}(r'_\star)}{{\cal W}_{\ell m}(r'_\star)}\, dr'_\star\quad , \quad {C}^{-} = \int_{r_\star}^{+\infty} \frac{\phi^{(1,0)(+)}_{\ell m}(r'_\star)\, S^{(1,0)}_{\ell m}(r'_\star)}{{\cal W}_{\ell m}(r'_\star)}\, dr'_\star\ ,\label{eq:Cpm}
\end{equation}
where ${\cal W}_{\ell m}$ is the constant Wronskian of the homogeneous 
solutions:
\begin{equation}
{\cal W}_{\ell m}(r_\star) = f\partial_r\phi^{(1,0)(+)}_{\ell m} \phi^{(1,0)(-)}_{\ell m} -
f\partial_r\phi^{(1,0)(-)}_{\ell m} \phi^{(1,0)(+)}_{\ell m}\ .
\end{equation}
The solution for $\phi^{(1,1)}_{\ell m}$ has the same form 
as Eq.~\eqref{eq:axialfullF}, with the substitution 
$S^{(1,0)}_{\ell m} \rightarrow S^{(1,1)}_{\ell m}$ in the $C^{\pm}$ 
coefficients.

For circular orbits the calculation of ${ C}^\pm$ greatly simplifies. 
In this case the source term can be written as function of Dirac's 
delta and it's first derivative:
\begin{equation}
S^{(1,0)}_{\ell m}=D(r,r_p)\delta(r-r_p)+G(r,r_p)\delta'(r-r_p)\ ,
\end{equation}
where $r_p$ is the secondary orbital radius, and the functions $D,G$ can 
be determined from the coefficients of $T^{p(1,0)}_{\mu\nu}$ (and 
of $T^{p(1,1)}_{\mu\nu}$ for the matter contribution). Integration 
in Eqs.~\eqref{eq:Cpm}can be performed analytically 
such that
\begin{equation}
{C}^+=\mathcal{C}^+\Theta(r-r_p)\quad \ , \quad {C}^-=\mathcal{C}^-\Theta(r_p-r)\ ,
\end{equation}
where 
\begin{equation}
\mathcal{C}^\pm=\frac{\phi^{(1,0)(\mp)}_{\ell m}(r_p)D(r_p)}{f_p{\cal W}}-
\frac{d}{dr}\left[\frac{\phi^{(1,0)(\mp)}_{\ell m}(r_p)G(r_p)}{{\cal W}f(r)}\right]_{r=r_p}\ .
\end{equation}
%

%
\subsection{$\ell \ge 2$ polar modes}
%

Perturbations in the polar sector are characterized by seven variables: 
four metric components \((H_{0,\ell m}, H_{1,\ell m}, H_{2,\ell m}, K_{\ell m})\), 
two components of the fluid velocity perturbation \((V^{(1,0)}_{\ell m}, W^{(1,0)}_{\ell m})\), 
and the density perturbation \(\rho^{(1,1)}\). Despite this  complexity, the 
dimensionality of the system can be significantly reduced.

The \(\theta\phi\) component of Eqs.~\eqref{eq:fields10} allows 
us to express \(H_{2,\ell m}\) in terms of \(H_{0,\ell m}\). 
Furthermore, the \(rr\), \(tr\), and \(t\theta\) components of 
Einstein equations can be used to eliminate the time derivative 
of \(H_{0,\ell m}\), yielding two coupled differential equations\footnote{These algebraic 
manipulations also introduce 
a third-order time derivative of \(K_{\ell m}\), which can be 
removed using the \(t\phi\) component of Einstein equations.}
that depend only on the metric functions \(K_{\ell m}\) and 
\(H_{1,\ell m}\), along with the fluid perturbations, and take the following form:

\begin{align}
&(b_1+b_2 \partial_r) H_{1,\ell m}+
(b_3 \partial_t+
b_4 \partial^2_{tr}) K_{\ell m}+b_5 V^{(1,0)}_{\ell m}+b_6 W^{(1,0)}_{\ell m}=S_{\ell m}^H\ \label{eq:masterZ1},\\
&(c_1+c_2\partial_{r}+c_3\partial^2_{rr}+c_4\partial^2_{tt})H_{1,\ell m}
+(c_4\partial_{t}+c_5\partial^2_{tr})K_{\ell m}+(c_6+c_7\partial_{r})V^{(1,0)}_{\ell m}+
c_8 W^{(1,0)}_{\ell m}=S_{\ell m}^K\ ,\label{eq:masterZ2}
\end{align}
From the time component of the covariant derivative of the stress-energy tensor, 
we obtain an equation for \(\rho^{(1,1)}_{\ell m}\):
\begin{equation}
d_1\partial_t \rho^{(1,1)}_{\ell m}
+d_2\partial_t K_{\ell m}
+(d_3+d_4\partial_r)H_{1,\ell m}+(d_5+d_6\partial_r) W^{(1,0)}_{\ell m}
+d_7V^{(1,0)}_{\ell m}=\mathcal{J}_{\ell m}^{\rho}\ .\label{eq:density1}
\end{equation}
The coefficients \((b_i,c_i,d_i)\) appearing in 
Eqs.~\eqref{eq:masterZ1}–\eqref{eq:density1} 
contain background quantities and depend only on 
\(r\). 
Solving Eq.~\eqref{eq:density1} allows to 
determine $\rho_{\ell m}^{(1,1)}$, and 
hence $(p_{r,\ell m}^{(1,1)}$,\ $p_{t,\ell m}^{(1,1)})$, 
as a function of background quantities and of 
the vacuum solution through Eq.~\eqref{eq:csdef}.

Finally, from \(\nabla_\mu T^{\mu\theta}=0\) 
and \(\nabla_\mu T^{\mu r}=0\), 
we obtain two first-order equations in time for 
\(\partial_t V_{\ell m}^{(1,0)}\) and \(\partial_t W_{\ell m}^{(1,0)}\).

We now reduce the coupled system for \(H_{1,\ell m}\) and \(K_{\ell m}\) 
to a single master equation for the metric perturbation, following 
the strategy introduced by Zerilli \cite{Zerilli:1970se, Zerilli:1971wd}, and isolate its \((0,1)\) and 
\((1,1)\) components. We first introduce the 
new functions \(\bar{\chi}_{\ell m}(t,r)\) 
and \(\bar{R}_{\ell m}(t,r)\):
\begin{align}
\partial_t K_{\ell m}(t,r)=\alpha \bar{\chi}_{\ell m}(t,r)+ \beta \bar{R}_{\ell m}(t,r)\quad\ ,\quad 
H_{1,\ell m}(t,r)=\gamma \bar{\chi}_{\ell m}(t,r)+ \delta \bar{R}_{\ell m}(t,r)\ ,\label{eq:zeriliicoeff1}
\end{align}
As in the axial sector, we introduce the scaling function 
\(Z(r)\) such that \(\chi_{\ell m}=\sqrt{Z}\bar{\chi}_{\ell m}\) and 
\(R_{\ell m}=\sqrt{Z}\bar{R}_{\ell m}\). 
The coefficients \((\alpha,\beta,\gamma,\delta)\), which depend 
only on \(r\), are fixed by requiring that \(\chi_{\ell m}\) 
and \(R_{\ell m}\) satisfy Zerilli-like equations of the form:
\begin{equation}
f\partial_r[f\partial_r\chi_{\ell m}]+({\cal V}^{P}-\partial_t^2)\chi_{\ell m}={\cal S}^{P}_{\ell m}\quad\ ,\quad
f \partial_r \chi_{\ell m}-R_{\ell m}={\cal J}^{P}_{\ell m}\ , \label{eq:polar01}
\end{equation}
for some scattering potential \({\cal V}^P\). At this stage, and for readability, 
we collectively include in the source terms \({\cal S}^{P}_{\ell m}\) 
and \({\cal J}^{P}_{\ell m}\) all contributions proportional 
to the secondary orbital configuration and fluid perturbations. Their explicit forms will be given later.

The coefficients that `diagonalize' the problem coincide with those 
originally found by Zerilli \cite{Zerilli:1970se, Zerilli:1971wd}. At this point, 
all metric perturbations can be expanded in the two-parameter scheme, 
e.g., \(\chi_{\ell m}=\chi_{\ell m}^{(0,1)}+\chi_{\ell m}^{(1,1)}\). 
As a result, Eqs.~\eqref{eq:masterZ1}–\eqref{eq:masterZ2} reduce to:
\begin{align}
\partial^2_{r_\star}\chi^{(1,0)}_{\ell m}&+(V^{P}-\partial_t^2)\chi^{(1,0)}_{\ell m}=S^{P^{(1,0)}}_{\ell m}\ ,\label{eq:polar10}\\
\partial^2_{r_\star}\chi^{(1,1)}_{\ell m}&+(V^{P}-\partial_t^2)\chi^{(1,1)}_{\ell m}
+(z_1+z_2f\partial_r)V^{(1,0)}_{\ell m}+(z_3+z_4f\partial_r)W^{(1,0)}_{\ell m}=S^{P^{(1,1)}}_{\ell m}\ .\label{eq:polar11}
\end{align}
The scattering potential for both the \((1,0)\) and \((1,1)\) equations 
coincides and is given by the well-known vacuum result:
\begin{align}
V^P=-\frac{2f}{r^3}\frac{9 M^3+9 \Lambda  M^2 r+3 \Lambda ^2 M r^2+\Lambda ^2 (\Lambda +1) r^3}{(3M+r\Lambda)^2}.
\end{align}
The source term \(S^{P^{(1,1)}}_{\ell m}\) is proportional to 
\(\chi_{\ell m}^{(1,0)}\) and to the components of \(T_{\mu\nu}^{p,(1,1)}\),  
while the coefficients \(z_{1,2,3,4}\) depend only on the background 
pressure and density. The density perturbation enters the equation 
for \(\chi^{(1,1)}_{\ell m}\) via the fluid velocities, which are determined by:
\begin{align}
\kappa_t\partial_t V^{(1,0)}_{\ell m}&+4\pi c^2_{t,{\ell m}}\rho_{\ell m}^{(1,1)}=S^V_{\ell m}\ ,\label{eq:masterV}\\
\kappa_r\partial_t W^{(1,0)}_{\ell m}&+(w_{1}+w_{2}f\partial_r)\rho_{\ell m}^{(1,1)}=S^W_{\ell m}\ ,\label{eq:masterW}
\end{align}
where \(\kappa_t=\rho(r)+p_t(r)\) and \(\kappa_r=\rho(r)+p_r(r)\). 
Finally, using Eqs.~\eqref{eq:polar10} and \eqref{eq:masterV}–\eqref{eq:masterW}, 
we can simplify the master equation for \(\rho_{\ell m}^{(1,1)}\). 
Taking the time derivative of Eq.~\eqref{eq:density1} yields:
\begin{align}\
(\partial_{r_\star}^2-c^{-2}_{r,{\ell m}}\partial_t^2+V^{\rho}
+\gamma_1\partial_{r_\star})\rho_{\ell m}^{(1,1)}=S_{\ell m}^{\rho}\ .\label{eq:masterrho} 
\end{align}
The coefficients \((w_1,w_2)\) involve combinations of \(p_{t,r}(r)\) 
and \(\rho(r)\), while \(V^\rho\) and \(\gamma_1\) depend only on the sound speeds. 
Along with the particle motion, the sources \(S^{V,W,\rho}_{\ell m}\) 
depend on the vacuum solutions \(\chi^{(1,0)}_{\ell m}\).

Note that Eq.~\eqref{eq:masterrho} is decoupled from the \((1,1)\) metric perturbations 
and can be solved once the vacuum solution is known. This allows to determine  
\(V^{(1,0)}_{\ell m}\) and \(W^{(1,0)}_{\ell m}\) via Eqs.~\eqref{eq:masterV}-\eqref{eq:masterW}. 
These quantities can then be used to fully solve the polar sector and obtain 
\(\chi_{\ell m}^{(1,1)}\) through Eq.~\eqref{eq:polar11}. The metric components can 
subsequently be reconstructed using the expressions in Appendix~\ref{app:metricrecon}.\\

We also briefly comment on the structure of the polar sector in the frequency domain. 
Although the equations remain too lengthy to present explicitly, the formulation 
simplifies significantly. In this case, the velocity perturbations, given by 
Eqs.~\eqref{eq:masterV}-\eqref{eq:masterW}, reduce to algebraic relations 
and can be eliminated from the wave equation for \(\chi^{(1,1)}_{\ell m}\), which 
can then be determined once a solution for \(\rho_{\ell m}^{(1,1)}\) is obtained using 
the Green function approach already discussed for the axial sector.

%
\subsection{$\ell =0$ modes}
%

For the sake of completeness, we complement the previous calculations
with the treatment of the $\ell = 0$ and $\ell = 1$ modes, which 
do not contribute to gravitational radiation.

Fo $\ell = m = 0$, only polar perturbations are excited. 
In this case we adopt the so called Zerilli gauge, which allows us to set 
$H_{1,00} = K_{00} = 0$~\cite{Zerilli:1971wd,Detweiler:2003ci}. 
Decomposing the remaining metric functions $H_{2,00}$ and $H_{0,00}$ 
into vacuum and matter components, we obtain
\begin{align}
\partial_{r}H^{(1,0)}_{0,00} &= -\frac{H^{(1,0)}_{2,00}}{rf} - 8\pi r A^{(1,0)}_{00}\ ,\\
\partial_{r}H^{(1,0)}_{2,00} &= -\frac{H^{(1,0)}_{2,00}}{rf} + \frac{8\pi r}{f^2} A^{0(1,0)}_{00}\ ,
\end{align}
which coincide with the standard results derived in the vacuum 
case~\cite{Sago:2021iku}, and
\begin{align}
\partial_r H_{0,00}^{(1,1)} &= -\frac{8 \pi  r c_r^2 \rho^{(1,1)}_{00}}{f}
-\frac{2}{f^2 r^2} \left(4 \pi  f r^3 p_r + m\right) H_{2,00}^{(1,0)} -\frac{H_{2,00}^{(1,1)}}{f r} - 8 \pi  r \mathcal{A}^{(1,1)}_{00}\ ,\\
\partial_r H_{2,00}^{(1,1)} &= -\frac{H_{2,00}^{(1,1)}}{f r}
+ \frac{8 \pi  r }{f} \rho^{(1,1)}_{00}
+ \frac{8 \pi r}{f^2} \mathcal{A}^{(0)(1,1)}_{00}  + \frac{2}{f^2 r^2} H_{2,00}^{(1,0)} \left(4 \pi  f r^3 \rho - m\right) \nonumber \\
&\quad - \frac{8 \pi}{f^3}  \mathcal{A}^{0(1,0)}_{00} (f r H - 2 m)\ .
\end{align}

Moreover, an algebraic equation for $W^{(1,0)}_{00}$ can be obtained from 
the $tr$ component of Einstein equations:
\begin{equation}
\kappa_r W^{(1,0)}_{00} =
\frac{2 i \sqrt{2} \pi \mathcal{A}^{1(1,1)}_{00}}{f}- \frac{\partial_t H_{2,00}^{(1,1)}}{2 f r}\ ,\label{eq:l=0-masterW}
\end{equation}
Finally, substituting the above into the $\theta\theta$ 
component of Eqs.~(\ref{eq:fields10}), we obtain a master 
equation for $\rho_{00}^{(1,1)}$:
\begin{align}
\left(\partial_{r_\star}^2 - c_{r,00}^{-2} \partial_t^2 + V^{\rho}_{\ell=0}
+ \gamma_{1,\ell=0} \partial_{r_\star} \right) \rho_{00}^{(1,1)} = S_{00}^{\rho}\ . \label{eq:l=0-masterrho}
\end{align}
The source term $S^{\rho}_{00}$ depends on the vacuum 
solution $H^{(1,0)}_{2,00}$ and on the secondary orbital 
trajectory, while the potential $V^p_{\ell=0}$ and the 
coefficient $\gamma_{1,\ell=0}$ contain terms proportional 
to the radial sound speed. As for the $\ell \geq 2$ modes, 
Eq.~\eqref{eq:l=0-masterrho} is decoupled from the $(1,1)$ 
metric perturbations and is entirely determined by the 
vacuum component. Once solved, one can determine $W^{(1,0)}_{00}$ 
via Eq.~\eqref{eq:l=0-masterW}, and subsequently reconstruct 
$H_{0,00}$ and $H_{2,00}$.

%
\subsection{$\ell =1$ modes}
%

For $\ell = 1$, both axial and polar modes are present. In the axial sector, the Zerilli gauge is implemented by setting $h_{0,1m} = 0$, leaving $h_{1,1m}$ as the only nonvanishing metric component to be determined \cite{Zerilli:1970se}. The field equations for the $(1,0)$ axial perturbation take the form:
\begin{align}
\partial_t^2 h^{(1,0)}_{1,1m} = &-rf\, 8i\pi\, \mathcal{Q}^{(1,0)}_{1m}\ , \\
\frac{2}{r^2} \partial_t h^{(1,0)}_{1,1m} &+ \frac{1}{r} \partial^2_{t,r} h^{(1,0)}_{1,1m} + \frac{8\pi}{f} \mathcal{Q}^{0(1,0)}_{1m} = 0\ .
\end{align}
At the $(1,1)$ order we have for the metric perturbation
\begin{align}
\partial^2_t h_{1,1m}^{(1,1)}&-[H\partial^2_t
-16\pi f (p_t-p_r)]h^{(1,0)}_{1,1m}+8\pi i f r \mathcal{Q}^{(1,1)}_{1m}=0\ ,\\
\big(f\partial^2_{t,r}&+\frac{2f}{r} \partial_t \big)h^{(1,1)}_{1,1m} +4f\kappa_t U^{(1,0)}_{1m}+8\pi r \mathcal{Q}_{1m}^{(0)(1,1)}-\frac{4}{r^2}\big(\pi r^3\kappa_r+m +\frac{rm}{2}\partial_r\big)\partial_t h^{(1,0)}_{1,1m}=0\ .
\end{align}
Finally, an equation for $\partial_t U^{(1,0)}_{1m}$ can be 
derived from $\theta$ component of the covariant divergence 
$\nabla_\mu T^{\mu \theta} = 0$. 

In the polar sector, we fix the Zerilli gauge by setting 
$K_{1m} = 0$, so that the remaining metric perturbations 
to determine are $H_{0,1m}$, $H_{1,1m}$, and $H_{2,1m}$, 
along with the fluid variables $V^{(1,0)}_{1m}$, $W^{(1,0)}_{1m}$, 
and $\rho^{(1,1)}_{1m}$. Decomposing the metric into its 
$(1,0)$ and $(1,1)$ components, we derive the corresponding 
perturbation equations by applying Einstein equations 
together with the $t$, $r$, and $\theta$ components of 
$\nabla_\mu T^{\mu \nu} = 0$. The  $(1,0)$ vacuum equations 
for $H_{0,1m}$, $H_{1,1m}$, and $H_{2,1m}$ coincide with 
those available in Appendix~B of \cite{Sago:2002fe}.

The functional forms of the equations for matter 
perturbations are identical to those in 
Eqs.~\eqref{eq:masterV}-\eqref{eq:masterrho}, valid 
for modes with $\ell \geq 2$, except for the
coefficients $w_1$, $w_2$, and $\gamma_1$, as well as 
the scattering potential $V^\rho$, whose explicit expressions 
are provided in the accompanying \texttt{Mathematica} file.

%
\section{Gravitational wave fluxes}
%

Having determined the axial and polar perturbations, 
we can compute the associated GW fluxes at infinity and 
at the horizon. The asymptotic structure of our 
metric allows us to employ the standard vacuum procedure 
\cite{Martel:2003jj,Martel:2005ir}, which relies on 
expressing the perturbations in a coordinate system 
where the metric exhibits the correct radial falloff 
\cite{Misner:1973prb}.

We note that matter fluxes across the horizon or 
to infinity are absent in this model. The perturbations 
of the matter stress–energy tensor are proportional 
to the background density, which vanishes at the 
horizon, and the fluid has compact support (or 
becomes rapidly negligible) at large radius. 
Consequently, no fluid perturbation can carry energy 
or angular momentum across either boundary. The 
secondary does excite fluid perturbations, but 
these remain confined within the matter distribution. 
The only radiative degrees of freedom at leading 
order are the standard GW modes, which carry imprints 
of matter through coupling and background effects. While no asymptotic matter fluxes are present, one may still expect local interactions between the perturber and the fluid. A Newtonian estimate suggests that a local drag force on the worldline — analogous to dynamical friction \cite{1943ApJ....97..255C} — would appear at 
order $\mathcal{O}(q^{2}\epsilon)$, sharing the same radiative scaling as the flux corrections discussed in the next section. Evaluating this effect would require computing the 
self-consistent motion of the secondary 
in the perturbed geometry, i.e. feeding 
the metric corrections back into the worldline evolution, a SF analysis that 
lies beyond the scope of this work.

To move from the RWZ gauge to the radiation gauge, 
we perform an infinitesimal coordinate transformation 
such that  
\begin{equation}
\delta g_{\mu\nu}^{\rm RG} = \delta g_{\mu\nu}^{\rm RWZ} - 
\nabla_{\mu} \xi_{\nu} - \nabla_{\nu} \xi_{\mu}\ ,\label{eq:gauge}
\end{equation}
where $\xi^\mu$ is a gauge vector expanded in  
multipole components (summation over $(\ell, m)$ is 
implicit):  
\begin{align}
\xi_\mu = (\alpha_1, \alpha_2, r^2[\alpha_3 \csc\theta\, &\partial_\phi + \alpha_4 \partial_\theta], r^2[\alpha_4 \partial_\phi - \alpha_3 \sin\theta\, \partial_\theta])\, Y_{\ell m}\ ,
\end{align}
with $\alpha_{1,2,3,4}$ being gauge functions dependent 
on $(t, r)$. Following~\cite{Martel:2003jj}, at infinity 
the perturbation tensor satisfies the outgoing 
radiation conditions: 
\begin{equation}
    \delta g_{\mu\nu}^{\rm ORG} n^\mu n^\nu = 
\delta g_{\mu\nu}^{\rm ORG} n^\mu m^\nu = 
\delta g_{\mu\nu}^{\rm ORG} n^\mu m^{\nu\ast}= \delta g_{\mu\nu}^{\rm ORG} n^\mu l^\nu = 
\delta g_{\mu\nu}^{\rm ORG} m^\mu m^{\nu\ast} = 0\ ,\label{eq:gaugeI}
\end{equation}
where the null tetrad $(l^\mu, n^\mu, m^\mu, m^{\mu\ast})$  
has components:
\begin{align}
l_\mu &= \left\{ -\sqrt{-g_{tt}g_{rr}}, g_{rr}, 0, 0 \right\}, \nonumber\\
n_\mu &= -\frac{1}{2} \left\{ \sqrt{-g_{tt}/g_{rr}}, 1, 0, 0 \right\}, \nonumber\\
m_\mu &= \frac{1}{\sqrt{2}} \left\{ 0, 0, r, i r \sin\theta \right\},
\end{align}
with $l^\mu l_\mu = n^\mu n_\mu = m^\mu m_\mu = m^{\mu\ast}m^{\ast}_{\mu} = 0$,  
$l^\mu n_\mu = -1 = -m^\mu m^{\ast}_\mu$, and the asterisk denoting complex conjugation~\cite{Chrzanowski:1975wv}.  

Equations~\eqref{eq:gaugeI}, together with the gauge transformation~\eqref{eq:gauge},  can be used to express 
$\alpha_{1,2,3,4}$ in terms of the RWZ metric perturbations  
and reconstruct the perturbation tensor at infinity.\footnote{These calculations  
are nearly identical to those in Appendix~B of~\cite{Martel:2003jj}, except for  
the general form of the metric components 
$g_{tt}$ and $g_{rr}$, which include matter 
contributions beyond Schwarzschild.}  

From the asymptotic form of the polar and axial components at $r \to \infty$,  
using Eqs.~\eqref{eq:h1rec}–\eqref{eq:Krec}, we find to leading order:
\begin{align}
h_{0\ell m} &\simeq -h_{1\ell m} \simeq r(\phi^{(1,0)}_{\ell m} + \phi^{(1,1)}_{\ell m})\ ,\\
H_{2,\ell m} &\simeq H_{0,\ell m} \simeq -H_{1,\ell m}\simeq r \partial_t (\chi_{\ell m}^{(1,0)} + \chi_{\ell m}^{(1,1)}) - \frac{2rM_e}{\Lambda} \partial_t^2 \chi_{\ell m}^{(1,0)}\ ,\\
K_{\ell m} &\simeq -(\chi_{\ell m}^{(1,0)} + \chi_{\ell m}^{(1,1)}) + \frac{2M_e}{\Lambda} \partial_t \chi_{\ell m}^{(1,0)}\ ,
\end{align}
assuming that at spatial infinity $\partial_t = -\partial_r + \mathcal{O}(1/r)$.
In the radiation zone, the perturbation becomes:
\begin{align}
\delta g_{AB}^{\rm ORG} = -2r^2 (\alpha_4 V^{\ell m}_{AB} + \alpha_3 W^{\ell m}_{AB}) + \mathcal{O}(1)\ ,\label{eq:hIRG}
\end{align}
where indices $A, B$ span the angular coordinates $(\theta, \phi)$, and
\begin{align}
\alpha_3 &= \frac{1}{r} \int^t (\phi^{(1,0)}_{\ell m} + \phi^{(1,1)}_{\ell m}) dt'\ ,\label{eq:a4IRG}\\
\alpha_4 &= -\frac{1}{2r} \int^t \left( \chi_{\ell m}^{(1,0)} + \chi_{\ell m}^{(1,1)} 
- \frac{2M_e}{\Lambda} \partial_t \chi_{\ell m}^{(1,0)} \right) dt'\ ,\label{eq:a3IRG}
\end{align}
and
\begin{equation}
V_{AB} =\left(\nabla_A \nabla_B  +\frac{\lambda}{2} \Omega_{AB} \right) Y_{\ell m}\quad\ , \quad\
W_{AB} = \frac{1}{2} \left[ \nabla_B \epsilon_A^{\ C} \nabla_C 
+ \nabla_A \epsilon_B^{\ C} \nabla_C \right] Y_{\ell m}\ ,
\end{equation}
with $\Omega_{AB} = \mathrm{diag}(1, \sin^2\theta)$,  
$\nabla_A$ the covariant derivative, and $\epsilon_{AB}$  
the Levi-Civita tensor on the unit 2-sphere.  

The energy and angular momentum fluxes can be obtained from the Isaacson stress-energy tensor 
for gravitational waves,  
\begin{equation}
T_{\mu\nu}^{\rm GW}=\frac{1}{64\pi}\langle\nabla_{\mu}\delta g^{\alpha\beta}\nabla_{\nu}\delta g_{\alpha\beta}\rangle\ ,\label{eq:TGW}
\end{equation}
where $\langle\ldots \rangle$ denotes average 
over a region of spacetime large compared with 
the GW wavelength. Given the symmetry of 
the background, we can express fluxes using 
the Killing vectors 
$\{\xi_{(t)}^{\nu},\xi_{(\phi)}^{\nu}\}$ 
associated to the two cyclic variables $t$ 
and $\phi$:
\begin{align}
-dE=&\int_{\Sigma}T^{{\rm GW}\mu}{_{\nu}}\xi_{(t)}^{\nu}d\Sigma_{\mu}=\pm\left[\frac{|g_{tt}|}{g_{rr}}\right]^{1/2} r^2\int_\Sigma T^{\rm GW}_{tr}d\Omega dt\ ,\label{eq:Eflux}\\
dL=&\int_{\Sigma}T^{{\rm GW}\mu}{_{\nu}}\xi_{(\phi)}^{\nu}d\Sigma_{\mu}= 
\pm\left[\frac{|g_{tt}|}{g_{rr}}\right]^{1/2} r^2\int_\Sigma T^{\rm GW}_{r\phi}d\Omega dt\ ,
\label{eq:Lflux}
\end{align}
where $d\Sigma_\mu$ is a surface element 
outward-oriented on $\Sigma$ and the signs $-$ and $+$ are for flux at horizon and at infinity respectively. Expanding all quantities 
at leading order 
in $1/r$, and using Eqs.~\eqref{eq:hIRG}–\eqref{eq:a4IRG} within the energy flux 
\eqref{eq:Eflux}, to order 
$\mathcal{O}(q^2 \epsilon)$ we obtain:
%
\begin{align}
\dot{E}^{\infty}_{\ell m} 
=& \frac{1}{64\pi} \frac{(\ell+2)!}{(\ell-2)!} \Bigg( 
\left| \chi^{(1,0)}_{\ell m} \right|^2 
+ 4 \left| \phi^{(1,0)}_{\ell m} \right|^2 + 2\, \mathrm{Re} \left[ \chi^{(1,0)}_{\ell m} \chi^{(1,1)\ast}_{\ell m} 
+ 4\phi^{(1,0)}_{\ell m} \phi^{(1,1)\ast}_{\ell m} \right.\nonumber\\
&\left.- \frac{2M_e}{\Lambda} \chi^{(1,0)}_{\ell m} \partial_t \chi^{(1,0)\ast}_{\ell m} \right]
\Bigg)\ .\label{eq:EInfflu}
\end{align}
%
Similarly, for the angular momentum flux, 
Eq.~\eqref{eq:Lflux}, we have: 
%
\begin{align}
\dot{L}^{\infty}_{\ell m} =&\frac{im}{128\pi} \frac{(\ell+2)!}{(\ell-2)!} \Bigg[ \chi^{(1,0)}_{\ell m} \int^t  \chi^{(1,0)\ast}_{\ell m} dt'
+ 4 \phi^{(1,0)}_{\ell m} \int^t dt'\phi^{(1,0)\ast}_{\ell m}- \chi^{(1,0)}_{\ell m} \left( \frac{2M_e}{\Lambda} \chi^{(1,0)\ast}_{\ell m} \right.\nonumber\\
&\left.- \int^t dt' \chi^{(1,1)\ast}_{\ell m} \right) -  \left( \frac{2M_e}{\Lambda}\partial_t \chi^{(1,0)}_{\ell m} 
-  \chi^{(1,1)}_{\ell m} \right)\int^t dt' \chi^{*(1,0)}_{\ell m} \nonumber\\
&+ 4 \phi^{(1,0)}_{\ell m} \int^t dt' \phi^{(1,1)\ast}_{\ell m}+ 4 \phi^{(1,1)}_{\ell m} \int^t dt' \phi^{(1,0)\ast}_{\ell m} 
\Bigg] + \mathrm{c.c.}\ .\label{eq:KInfflu}
\end{align}
%
The first two terms in Eqs.~\eqref{eq:EInfflu} and~\eqref{eq:KInfflu}  
correspond to the standard fluxes at infinity  
for vacuum perturbations around Schwarzschild BHs.\\

Calculations of GW fluxes at the horizon 
proceed analogously to those at infinity. We 
impose an ingoing radiation gauge by swapping 
$l^\mu \leftrightarrow n^\mu$ in Eqs.~\eqref{eq:gaugeI}, and express the gauge functions in terms of the RWZ metric perturbations near the horizon, i.e., in the limit 
$f \rightarrow 0$.

Using Eqs.~\eqref{eq:h1rec}–\eqref{eq:Krec}, we obtain the leading-order behavior of the axial 
and polar components as $r \to 2M$:
\begin{align}
f h_{1\ell m} &\simeq - 2M\left(\phi^{(1,0)}_{\ell m} + \phi^{(1,1)}_{\ell m}\right) + \frac{3MH_h}{2} \phi^{(1,0)}_{\ell m}\ , \\
h_{0\ell m} &\simeq - 2M\left(\phi^{(1,0)}_{\ell m} + \phi^{(1,1)}_{\ell m}\right) - \frac{M}{2} H_h \phi^{(1,0)}_{\ell m}\ , \\
H_{2,\ell m} &\simeq H_{0,\ell m} \simeq H_{1\ell m} \simeq \frac{1}{2}(4M\partial_t - 1)\left[\chi_{\ell m}^{(1,0)} + \chi_{\ell m}^{(1,1)}\right]- \frac{H_h}{8}(4M\partial_t - 1)\chi_{\ell m}^{(1,0)}\ , \\
\partial_t K_{\ell m} &\simeq \left(\frac{\Lambda + 1}{2M} + \partial_t\right)\left(\chi_{\ell m}^{(1,0)} + \chi_{\ell m}^{(1,1)}\right) - \frac{H_h}{4} \left(\frac{\Lambda + 1}{2M} + \partial_t\right) \chi_{\ell m}^{(1,0)}\ ,
\end{align}
where $H_h \equiv H(r = r_h)$, and we assume 
that near the horizon $\partial_t = f \partial_r + \mathcal{O}(f)$. Combining\footnote{Following \cite{Martel:2003jj} 
we rescale $\alpha_2\rightarrow \alpha_2f^{-1}(1-H_h)$.} these expressions with 
Eqs.~\eqref{eq:gauge} and \eqref{eq:gaugeI}, we 
can write the metric perturbation in the 
ingoing radiation gauge as:
\begin{align}
\delta g_{AB}^{\rm IRG} = -8M^2 \left(\alpha_4 V^{\ell m}_{AB} + \alpha_3 W^{\ell m}_{AB}\right) + \mathcal{O}(f)\ , \label{eq:hHRG}
\end{align}
with the gauge coefficients given by
\begin{align}
\alpha_3 &= -\frac{1}{2M} \int^t \left(\phi_{\ell m}^{(1,0)} + \phi_{\ell m}^{(1,1)} - \frac{H_h}{4} \phi^{(1,0)}_{\ell m} \right) dt'\ , \label{eq:a3HRG} \\
\alpha_4 &= -\frac{1}{4M} \int^t \left(\chi^{(1,0)}_{\ell m} + \chi^{(1,1)}_{\ell m} + \frac{MH_h}{3 + 2\Lambda} \partial_t \chi^{(1,0)}_{\ell m} - \frac{H_h}{4} \chi^{(1,0)}_{\ell m} \right) dt'\ . \label{eq:a4HRG}
\end{align}
The calculation of energy and angular momentum 
fluxes proceeds similarly to the far-zone treatment~\cite{Martel:2003jj}, by isolating the $\mathcal{O}(f^{-1})$ contribution to the GW 
stress-energy tensor \eqref{eq:TGW}, and 
neglecting terms of order $\mathcal{O}(1)$. 

We substitute the expression of the metric 
perturbation \eqref{eq:hHRG} into Eqs.~\eqref{eq:Eflux}-\eqref{eq:Lflux}, 
also multiplying by a $-$ sign to account 
that we compute BH absorption rather fluxes 
in the radiation zone. To the leading order 
in $f$  we find:
%
\begin{align}
\dot{E}^{\rm H}_{\ell m} =& \frac{1}{64\pi} \frac{(\ell+2)!}{(\ell-2)!} \Bigg(\left| \chi^{(1,0)}_{\ell m} \right|^2 
+ 2\, \mathrm{Re} \left[ \chi^{(1,0)}_{\ell m} \chi^{(1,1)\ast}_{\ell m}  
+ \frac{MH_h}{3+2\Lambda} \chi^{(1,0)}_{\ell m} \partial_t \chi^{(1,0)\ast}_{\ell m} 
\right]
+ 4 \left| \phi^{(1,0)}_{\ell m} \right|^2 \nonumber\\ 
&+  8\, \mathrm{Re} \left[\phi^{(1,0)}_{\ell m} \phi^{(1,1)\ast}_{\ell m} 
\right]
\Bigg)\ .\label{eq:Ehorflu}\\
\dot{L}^{\rm H}_{\ell m} =&\frac{im}{128\pi} \frac{(\ell+2)!}{(\ell-2)!} \Bigg[ \chi^{(1,0)}_{\ell m} \int^t  \chi^{(1,0)\ast}_{\ell m} dt'
+ 4 \phi^{(1,0)}_{\ell m} \int^t dt'\phi^{(1,0)\ast}_{\ell m}
\nonumber\\ 
&+\chi^{(1,0)}_{\ell m} \int^t dt'\left( \frac{MH_h}{3+2\Lambda} \partial_t\chi^{(1,0)\ast}_{\ell m} 
+  \chi^{(1,1)\ast}_{\ell m} \right)+\left( \frac{MH_h}{3+2\Lambda} \partial_t\chi^{(1,0)}_{\ell m} 
+  \chi^{(1,1)}_{\ell m} \right) \int^t dt' \chi^{(1,0)*}_{\ell m}\nonumber\\ 
& + 4 \phi^{(1,0)}_{\ell m} \int^t dt' \phi^{(1,1)\ast}_{\ell m} 
 + 4 \phi^{(1,1)}_{\ell m} 
 \int^t dt' \phi^{(1,0)\ast}_{\ell m}
\Bigg] + \mathrm{c.c.}\ .\label{eq:Lhorflu}
\end{align}
%
The first two terms in Eqs.~\eqref{eq:Ehorflu}–\eqref{eq:Lhorflu} represent vacuum contributions to the energy and angular momentum fluxes. The remaining terms depend on the matter distribution and vanish in the limit $\epsilon \rightarrow 0$.

%
\section{Conclusions}
%

In this work, we developed a multi-parameter framework to 
model the dynamics and GW emission of binaries with large 
mass asymmetries embedded in dense astrophysical environments. 
Previous studies have emphasized the scientific potential 
of such systems to probe the properties of baryonic and 
dark matter evolving alongside compact objects \cite{Cardoso:2022whc,Speeney:2024mas,Gliorio:2025cbh}. 
However, these efforts also highlighted the significant 
complications introduced by non-vacuum environments, 
which have so far made accurate waveform modeling intractable.

Motivated by these challenges, we constructed a semi-analytical 
approach that treats matter effects as small perturbations to 
vacuum spacetime, as  supported by most realistic astrophysical 
scenarios. By expanding Einstein equations around the 
Schwarzschild solution in powers of the binary mass ratio 
and the ratio of environmental to BH density, we derived 
expressions for both metric and matter perturbations within 
a genuinely SF framework at adiabatic order.

Our key results, summarized in Eqs.~\eqref{eq:ax01F}–\eqref{eq:ax11F}, \eqref{eq:polar10}–\eqref{eq:polar11}, and \eqref{eq:masterV}–\eqref{eq:masterrho}, show that both axial and polar perturbations reduce to equations closely resembling the 
well-known Regge-Wheeler and Zerilli formalisms. Notably, 
unlike previous studies, we demonstrate that polar modes 
can be captured by a single Zerilli-like master variable, 
greatly simplifying numerical computations. We provide 
explicit expressions for reconstructing the metric functions 
and computing GW fluxes for binaries on generic orbits. 

This framework represents an initial step toward the development 
of accurate and computationally feasible waveform models for asymmetric binaries in complex environments — key targets for 
future GW detectors like LISA. It also offers a flexible tool 
to study the interaction of such systems with ambient matter 
via time-domain evolution, and to investigate properties typically 
studied in vacuum, such as BH quasinormal mode spectra \cite{Pezzella:2024tkf,Spieksma:2024voy}. However, several 
advancements are necessary to reach full astrophysical 
realism.

One major, yet essential, challenge lies in modeling binaries 
with a rotating primary. Describing matter perturbations 
around Kerr BHs could benefit from recent progress in 
modeling vacuum perturbations within modified gravity theories, 
assuming small deviations from GR 
\cite{Li:2022pcy,Hussain:2022ins,Cano:2023tmv,Cano:2023jbk}. 
In principle, the BH spin could be introduced as a third 
perturbative parameter within a slow-rotation scheme, such 
as the Hartle-Thorne formalism \cite{1968ApJ...153..807H}. 
However, this approach generally exhibits poor convergence 
at high spin values, which are expected for astrophysical 
BHs. The fluid description could also be enhanced in multiple 
ways, for instance by investigating the impact of viscous 
effects on the binary dynamics \cite{Boyanov:2024jge}.

While our focus here is methodological, and the 
present model still has limited direct astrophysical 
applicability, due to the absence of spin and the 
restriction to spherically symmetric matter topologies, 
it nonetheless provides a first consistent framework 
for matter-embedded compact binaries. Interestingly, 
spherically symmetric configurations of BHs immersed in 
dense gas could, in fact, be relevant to certain recently 
observed compact sources — although at high redshift — 
the so-called ``red dots,'' which may represent heavily 
enshrouded accreting BHs 
\cite{2025arXiv250821748J,Begelman:2025upi,Hassan:2025xac}.

Finally, current studies of the evolution of asymmetric 
binaries including radiation reaction have mostly been 
restricted to circular, equatorial orbits due to computational 
complexity (see Ref.~\cite{Duque:2024mfw} for a study on the 
relevance of eccentricity in binaries immersed in an accretion 
disk). The framework developed here allows exploration of EMRI 
and IMRI dynamics on generic, eccentric, and inclined orbits 
across a broad parameter space, and assessment of the impact of 
matter on parameter estimation using recent tools developed 
to analyze GW signals from asymmetric binaries 
\cite{Katz2021,Chapman-Bird:2025xtd,Cole:2025sqo}.

\section*{Acknowledgements}
We thank Rodrigo Vicente, Laura Sberna and Konstantinos 
Kritos for interesting and useful discussions. We also 
thank Vitor Cardoso, Nicholas Speeney, Richard Brito, 
Kyriakos Destounis for having carefully read this 
manuscript and for useful comments. We also thank the 
Referees for their valuable comments, which have helped 
improve the quality of this manuscript.

\paragraph{Funding information}
A.M. acknowledges financial support from MUR PRIN 
Grants No.~2022-Z9X4XS and No.~2020KB33TP. 
S.D. acknowledges financial support from MUR, PNRR - Missione 4 - Componente 2 - Investimento 1.2 - finanziato dall'Unione europea - NextGenerationEU (cod. id.: SOE2024\_0000167, CUP:D13C25000660001).

\begin{appendix}
\numberwithin{equation}{section}

%
\section{Metric perturbations as a function of the master variables}\label{app:metricrecon}
%
Metric perturbations can be easily reconstructed once a solution 
for the master equations \eqref{eq:ax01}-\eqref{eq:ax11} 
and \eqref{eq:polar10}-\eqref{eq:polar11} have been found. 
In this Appendix we provide relations that determine axial 
and polar metric functions at the linear order in $\mathcal{O}(\epsilon)$. 
In the Regge-Wheeler gauge for axial modes with 
$\ell \ge2$ we have:
\begin{align}
h_{1,\ell m}=&-r f^{-1}\phi^{(1,0)}_{\ell m}+
\frac{3[fr H-2m]}{4f^2}\phi^{(1,0)}_{\ell m}-rf^{-1}\phi^{(1,1)}_{\ell m}\ ,\label{eq:h1rec}\\
\partial_t h_{0,\ell m}=&-f\partial_r\left(r\phi^{(1,0)}_{\ell m}\right)
+f\frac{8i\sqrt{2}\pi r^2{\cal D}^{(1,0)}_{\ell m}}{\sqrt{\lambda(\lambda-2)}}+f\frac{8i\sqrt{2}\pi r^2[H{\cal D}^{(1,0)}_{\ell m}+{\cal D}^{(1,1)}_{\ell m}]}{\sqrt{\lambda(\lambda-2)}}
\nonumber\\
&-f\partial_r\left(r\phi^{(1,1)}_{\ell m}\right)+\frac{1}{4}\left(\frac{2m}{r}-fH\right)\partial_r\left(r\phi^{(1,0)}_{\ell m}\right)+\left[\frac{m}{fr}+2\pi r^2(p_r-\rho)\right]\phi^{(1,0)}_{\ell m}\,\label{eq:h0rec}
\end{align}
where $f=1-2M/r$ and $\lambda=\ell(\ell+1)$. Frequency domain expressions 
can be obtained by replacing time derivatives as $\partial_t\rightarrow -i\omega$.

The reconstruction of polar perturbations is more convoluted. 
We provide here explicit expressions including only the 
master functions. The full form depending on 
the coefficients of the secondary stress-energy tensor 
is provided in the \texttt{Mathematica} supplementary file:
\begin{align}
\partial_t H_{0,\ell m}=[A_1&+A_5+(A_2+A_6)\partial_r +A_3\partial_r^2 +A_{4}\partial^3_{r}]\chi_{\ell m}^{(1,0)}+(A_1+A_2 \partial_{r}+rf \partial^2_{r})\chi_{\ell m}^{(1,1)}\nonumber\\
&+(A_7+B_4\partial_r )V^{(1,0)}_{\ell m}+(A_8+B_5\partial_r)W^{(1,0)}_{\ell m}+S_{\ell m}^{H_0}\ ,\label{eq:H0rec}\\
H_{1,\ell m}=(B_1&+B_6+B_2\partial_r+B_3\partial_r^2) \chi_{\ell m}^{(1,0)}+(B_1+r\partial_r)\chi_{\ell m}^{(1,1)}+ \frac{B_4}{f} V_{\ell m}^{(1,0)}+\frac{B_5}{f} W^{(1,0)}_{\ell m}+S_{\ell m}^{H_1}\ ,\label{eq:H1rec}\\
\partial_t K_{\ell m}=\bigg(C_1&+C_4+\frac{f}{r}B_2\partial_r+\frac{f}{r}B_3\partial_r^2\bigg) \chi_{\ell m}^{(1,0)}+(C_1+f\partial_r) \chi_{\ell m}^{(1,1)}
+\frac{B_4}{r}V^{(1,0)}_{\ell m}+\frac{B_5}{r}W^{(1,0)}_{\ell m}+S_{\ell m}^{K}\ ,\label{eq:Krec}
\end{align}
where the source terms $S_{\ell m}^{H_0.H_1,K}$ 
depend on the particle orbital motion, and 
\begin{align}
A_1=&-\frac{9 M^3 + 9 M^2 r \Lambda + 3 M r^2 \Lambda^2 + r^3 \Lambda^2 (1 + \Lambda)}{r^2 {\cal C}^2}\quad\ ,\quad A_2=\frac{3 M^2 - M r \Lambda + r^2 \Lambda}{r {\cal C}}\ , \\ 
A_3=&r f-\frac{rf H}{4}-\frac{4 \pi  f r^4 (2p_r-3 \rho )}{{\cal C}}+\frac{m [(2-3 \Lambda ) r-13 M]}{2{\cal C}}\quad\ ,\quad A_{4} = -\frac{2 r^2f m}{{\cal C}}\ ,\\
A_5 =&-\frac{A_1H}{4}+\frac{m}{2r^4f^2{\cal C}^4}\bigg\{9 (14 \Lambda -3) M^5 r-54 M^6+3 [\Lambda  (124 \Lambda -33)+36] M^4 r^2\nonumber\\
&+60 (\Lambda -1) \Lambda  (2 \Lambda -3) M^3 r^3+3 \Lambda ^2 [2 \Lambda  (7 \Lambda -6)+55] M^2
r^4+\Lambda ^3[\Lambda  (2 \Lambda -33)-6] M r^5\nonumber\\
&+\Lambda ^3 [12-(\Lambda -9) \Lambda ] r^6\bigg\}
-\frac{2 \pi \rho}{f{\cal C}^3} \bigg\{18 M^4+9 (1-4 \Lambda ) M^3 r+6 \Lambda  (\Lambda +12) M^2 r^2\nonumber\\
&+\Lambda  [(3-4 \Lambda ) \Lambda -12] M r^3+4 \Lambda ^2 (\Lambda +1) r^4\bigg\}-\frac{4 \pi  r}{{\cal C}^2} \left[15 M^2+6 \Lambda  M r+\Lambda  (3 \Lambda +4) r^2\right]p_t\nonumber\\
&+\frac{2 \pi  r^2}{{\cal C}^2} \left[3 M^2+\Lambda  (\Lambda +2) r^2\right]\rho'-\frac{2 \pi}{f{\cal C}^3}  \bigg\{72 M^4+3 (34 \Lambda -15) M^3 r+6 \Lambda  (2 \Lambda -15) M^2 r^2\nonumber\\
&+\Lambda  [\Lambda  (6 \Lambda -5)+12] M r^3-4 \Lambda ^2 (\Lambda +1) r^4\bigg\}p_r\ ,\\
A_6 =&-\frac{A_2H}{4}+\frac{4 \pi  r^2 \rho}{{\cal C}^2} \left[(6-9 \Lambda ) M r+4 \Lambda  r^2-15 M^2\right]+\frac{4 \pi  r^2}{{\cal C}}[M-(\Lambda +2) r]p_r\nonumber\\
&-\frac{16 \pi  r^3f p_t}{{\cal C}}+\frac{4 \pi  r^4f \rho '}{{\cal C}}-\frac{m}{2 r^2f {\cal C}^3}\bigg\{9 M^4+(36-69 \Lambda ) M^3 r-9 (\Lambda -13) \Lambda  M^2 r^2\nonumber\\
&+\Lambda  [(14-11 \Lambda ) \Lambda -12] M r^3+\Lambda ^2 (9 \Lambda +8) r^4\bigg\}\ , \\
A_7 =&\frac{4f}{{\cal C}^2}[3 M^2 + r^2 \Lambda^2 + M r (2 \Lambda-3 )]\kappa_t -\frac{4 r^2f^2} {{\cal C}} \kappa'_t\ ,\\
A_8 =&\frac{2rf}{{\cal C}^2}\left\{rf{\cal C} (r\rho'-2p_t)-[9 M^2 + (5 M - r) r \Lambda] p_r + [3 M (2 r-M) + r (M + r) \Lambda] \rho \right\}\ ,\\
B_1 =&\frac{\Lambda  r}{{\cal C}}-\frac{M}{f r} \quad \ ,\quad  B_2 = r -\frac{rH}{4} -\frac{m}{2f{\cal C}^2}[3M^2 + 6 M r (1 + \Lambda) - r^2 \Lambda (2 + \Lambda)]   \ , \\
&+ \frac{4 \pi r^4(\rho-2 p_r)}{{\cal C}}
\quad\ ,\quad B_3 =\frac{2 r^2 m}{{\cal C}}\quad\ , \quad B_4 = -\frac{4 r^2f^2 \kappa_t}{{\cal C}}
\ ,\\
B_5 =& -\frac{2 f^2 r^3\kappa_r}{{\cal C}}\quad\ ,\quad B_6=\frac{2 \pi r^2}{f{\cal C}^2}\left[3 M^2+\Lambda  (\Lambda +2) r^2\right]\rho -\frac{H}{4}B_1\nonumber\\
&-\frac{3 m}{2r^2f^2{\cal C}^3}\bigg[33 M^4+(31 \Lambda +6) M^3 r+\Lambda ^2 (\Lambda +2) r^4+3 \Lambda  (5 \Lambda +3) M^2 r^2\nonumber\\
   &+\Lambda  \left(\Lambda ^2+2\right) M
   r^3\bigg]-\frac{2 \pi r^2}{f{\cal C}^2} \left[15 M^2+6 \Lambda  M r+\Lambda (3 \Lambda +4) r^2\right]p_r\ ,\\
C_1 =&\phantom{+}\frac{6 M^2 + 3 M r \Lambda + r^2 \Lambda (1 + \Lambda)}{r^2 {\cal C}}\quad \ ,\quad 
C_4 =-\frac{H}{4}C_1+\frac{2 \pi  r}{{\cal C}^2}\rho  \left[3 M^2+\Lambda 
   (\Lambda +2) r^2\right]\nonumber\\
&- \frac{m}{2
   f r^3 {\cal C}^3} \bigg[18 M^4-3 (5 \Lambda -6) M^3 r-9 (\Lambda -3) \Lambda  M^2 r^2-3 \Lambda  \left(3 \Lambda ^2-2\right) M r^3\nonumber\\
   &-\Lambda ^2 ((\Lambda -3) \Lambda -6) r^4\bigg]-\frac{2 \pi  r}{{\cal C}^2}\left[15 M^2+6 \Lambda  M r+\Lambda  (3 \Lambda +4) r^2\right]p_r\ ,
\end{align}
with $\Lambda=(\ell+2)(\ell-1)/2$, $\kappa_{t,r}=p_{t,r}+\rho$, 
$\mathcal{C}=r\Lambda+3M$
and a prime denoting radial derivative.

%
\section{Decoupling of Axial and Polar Modes into vacuum and 
matter components using the scaling function $Z$.}\label{app:scaling}
%

In this appendix, we clarify why, in computing axial and polar modes, we 
chose to work with a single metric perturbation rather than separating 
vacuum $(0,1)$ and matter $(1,1)$ components from the beginning.

The structure of the equations for axial modes, for example, allows one 
to follow a procedure similar to the vacuum case. In this framework, it 
is possible to eliminate one of the metric functions at each order in 
$\epsilon$ using the Einstein equations, leading to two second-order 
differential equations in $(r, t)$ for the $(1,0)$ and $(1,1)$ perturbations. 
These equations can then be recast in the familiar wave-like form by 
introducing a generalized tortoise coordinate, which facilitates the 
imposition of boundary conditions at spatial infinity and the BH horizon.
However, a subtlety arises from the fact that the generalized tortoise 
coordinate $dr^\star/dr = 1/\sqrt{-g_{tt}/g_{rr}}$, depends on the parameter 
$\epsilon$. This introduces an ambiguity due to the perturbative relation 
between $r$ and $r_\star$, since $dr^\star/dr = f^{-1} + \mathcal{O}(\epsilon)$, 
on whether one should use $r$ or $r_\star$ in the perturbative 
expansion (see \cite{Hatsuda:2023geo} for further details). This 
issue can be circumvented by following the approach developed in 
\cite{Cardoso:2019mqo}, which we briefly outline here.

Consider a scalar perturbation $\Phi$ on a fixed, spherically symmetric 
background with the metric:
\begin{equation}
ds^2 = -A(r)dt^2 + \frac{dr^2}{B(r)} + r^2(d\theta^2 + \sin^2\theta\, d\phi^2)\,.
\end{equation}
After decomposing $\Phi$ into spherical harmonics, the Klein--Gordon 
equation $\Box \Phi = 0$ can be written as:
\begin{equation}
-\frac{\partial^2 \Phi}{\partial t^2} +{\cal F} \frac{d}{dr}\left({\cal F} \frac{d\Phi}{dr}\right) - {\cal F} V \Phi = 0\,, \label{eq:appmaster1}
\end{equation}
where $V$ is the effective potential, which depends on the background 
geometry. Assume the metric functions $A(r)$ and $B(r)$ are close to 
the Schwarzschild solution:
\begin{equation}
A(r) = \left(1 - \frac{r_h}{r}\right)(1 + \delta A), \quad B(r) = \left(1 - \frac{r_h}{r}\right)(1 + \delta B)\ ,
\end{equation}
with $\delta A, \delta B \ll 1$, and where $r_h$ denotes the horizon 
radius.\footnote{Note that in general $r_h$ may differ from the Schwarzschild 
value. In such cases, $r_h$ should be treated as a fundamental parameter 
in the computation of perturbations, as done in the cases studied in 
\cite{Cardoso:2019mqo}.} Then, at the leading order in the metric changes 
$(\delta A, \delta B)$, the function ${\cal F} = \sqrt{AB}$ can be 
expressed as:
\begin{equation}
{\cal F} = f(r)Z(r) = \left(1 - \frac{r_h}{r}\right) Z(r) = \left(1 - \frac{r_h}{r} \right)[1 + \delta Z(r)]\,.\nonumber
\end{equation}
Introducing the rescaled field $\phi = \sqrt{Z} \Phi$, and expanding 
Eq.~\eqref{eq:appmaster1} to linear order in $\delta Z$, the master 
equation becomes:
\begin{equation}
-(1+2\delta Z)\frac{\partial^2 \phi}{\partial t^2} + f \frac{d}{dr} \left(f \frac{d\phi}{dr} \right) - f \tilde{V} \phi = 0\,, \label{eq:appmaster2}
\end{equation}
where \(\tilde{V}\) is the modified potential (the explicit form can be found 
in \cite{Cardoso:2019mqo}). For both axial and polar sectors, the metric 
perturbations we find satisfy master equations analogous to Eq.~\eqref{eq:appmaster1}, 
with $r_h = 2M$, and can be recast into the form of Eq.~\eqref{eq:appmaster2} 
by introducing an appropriate scaling function $Z$. Since the prefactor of 
the radial derivative terms in Eq.~\eqref{eq:appmaster2} is $f(r)$, we can 
adopt the standard tortoise coordinate $r^\star = r + 2M \ln\left(r/2M - 1\right)$. 
This allows us to write the perturbations as a sum of the $(1,0)$ and \((1,1)\) 
components, and isolate their contributions without introducing ambiguities.

%
\section{Coefficients of the particle stress-energy momentum tensor}\label{app:stressenergy}
%

The form of the coefficients $\{\mathcal{A}^{0(1,0)}_{\ell m},\ldots \mathcal{F}^{(1,0)}_{\ell m}\}$ 
of the particle stress-energy tensor, can be found by projecting 
each one of the ten tensor harmonics on Eq.~\eqref{eq:PP-EMT10}. 
Introducing the scalar product between two tensor harmonics $A_{\mu\nu}$ 
and $B_{\mu\nu}$: 
\begin{equation}
(A,B)=\int\int\eta^{\mu\sigma} \eta^{\nu\delta}A^\ast{\mu\nu}B_{\sigma\delta}\sin\theta d\theta d\phi\ ,
\end{equation}
where $\eta_{\mu\nu}$ is the Minkowski metric tensor in spherical coordinates, 
and $\ast$ denotes complex conjugation, we have, for example, 
${\cal A}^{(1,1)}_{\ell m}=({\bf a}_{\ell m}, T^p{}^{(1,1)})$. We provide the 
expression of the coefficients for generic orbits in the supplementary material.
In the case of equatorial circular motion, $\theta_p=\pi/2 $, for a secondary at 
a radius $r=r_p$, the only non vanishing coefficients are given by $Q^{0(1,0)}_{\ell m}$ 
for the axial sector, and $({\cal A}^{0(1,0)}_{\ell m},{\cal B}^{0(1,0)}_{\ell m},{\cal G}^{0(1,0)}_{\ell m},
{\cal D}^{0(1,0)}_{\ell m},{\cal F}^{0(1,0)}_{\ell m})$ for the polar modes (and 
similarly for the $(1,1)$ coefficients). Their explicit form is given by:
%
\begin{align}
{\cal Q}^{0(1,0)}=&\frac{\sqrt{2}f {\cal L}^{(0,0)}}{r^3\sqrt{\lambda}}\partial_\theta Y_{\ell m}^\ast\delta(r-r_p)\ \ ,\ \
{\cal A}^{0(1,0)}=\frac{f {\cal E}^{(0,0)}}{r^2}Y_{\ell m}^\ast\delta(r-r_p)\ , \\ 
{\cal B}^{0(1,0)}=&
\frac{i\sqrt{2}f{\cal L}^{(0,0)}}{r^3\sqrt{\lambda}}\partial_\phi Y_{\ell m}^\ast\delta(r-r_p)\ ,\\
{\cal D}^{(1,0)}=&\frac{i\sqrt{2}f({\cal L}^{(0,0)})^2}{r^4{\cal E}^{(0,0)}\sqrt{\lambda(\lambda-2)}}\partial_{\theta\phi}Y_{\ell m}^\ast\delta(r-r_p)\ ,\\
{\cal F}^{(1,0)}=&
\frac{f({\cal L}^{(0,0)})^2\delta(r-r_p)}{\sqrt{2}r^4{\cal E}^{(0,0)}\sqrt{\lambda(\lambda-2)}}
[\partial^2_{\phi}-\partial^2_{\theta}]
Y_{\ell m}^\ast
\quad \ , \quad 
{\cal G}^{(1,0)}=\frac{f({\cal L}^{(0,0)})^2}{\sqrt{2}r^4{\cal E}^{(0,0)}}
Y_{\ell m}^\ast\delta(r-r_p)\ ,\\
Q^{0(1,1)}_{\ell m}=&\frac{1}{\sqrt{2\lambda}r^4}[2 f r {\cal L}^{(0,1)}+f r {\cal L}^{(0,0)}H
-2{\cal L}^{(0,0)} m]\partial_\theta Y^\ast_{\ell m}\delta(r-r_p)\ ,\\
{\cal A}^{0(1,1)}_{\ell m}=&\frac{1}{2r^3}[2fr {\cal E}^{(0,1)}+
(rfH-2m){\cal E}^{(0,0)}]
Y^\ast_{\ell m}\delta(r-r_p)\ ,\\
{\cal B}^{0(1,1)}_{\ell m}=&
\frac{i}{\sqrt{2\lambda}r^4}[2fr {\cal L}^{(0,1)}
+fr{\cal L}^{(0,0)}H-2{\cal L}^{(0,0)}m]
\partial_\phi Y^\ast_{\ell m}\delta(r-r_p)\ ,\\
{\cal D}^{0(1,1)}_{\ell m}=&
\frac{i{\cal L}^{(0,0)}\delta(r-r_p)}{r^5({\cal E}^{(0,0)})^2\sqrt{2\lambda(\lambda-2)}}
\{4fr {\cal E}^{(0,0)}
{\cal L}^{(0,1)}
-2fr {\cal E}^{(0,1)}{\cal L}^{(0,0)}\nonumber\\
&+{\cal E}^{(0,0)}{\cal L}^{(0,0)}[frH-2m]\}
\partial_{\theta\phi}Y_{\ell m}^\ast\ ,\\
{\cal F}_{\ell m}^{(1,1)}
=&\frac{{\cal L}^{(0,0)}\delta(r-r_p)}{r^5({\cal E}^{(0,0)})^2\sqrt{8\lambda(\lambda-2)}}
\{4fr {\cal E}^{(0,0)}
{\cal L}^{(0,1)}
-2fr {\cal E}^{(0,1)}
{\cal L}^{(0,0)}
\nonumber\\
&+{\cal E}^{(0,0)}
{\cal L}^{(0,0)}[frH-2m]\}
[\partial^2_{\phi}-\partial^2_{\theta}]
Y_{\ell m}^\ast\ ,\\
{\cal G}_{\ell m}^{(1,1)}
=&
\frac{{\cal L}^{(0,0)}\delta(r-r_p)}{2\sqrt{2}r^5({\cal E}^{(0,0)})^2}
\{4fr {\cal E}^{(0,0)}
{\cal L}^{(0,1)}
-2fr {\cal E}^{(0,1)}
{\cal L}^{(0,0)}
+{\cal E}^{(0,0)}
{\cal L}^{(0,0)}[frH-2m]\}
Y_{\ell m}^\ast\ ,
\end{align}
%
where spherical harmonics are evaluated at $\theta=\theta_p(t)$ and $\phi=\phi_p(t)$, 
while ${\cal E}^{(0,1)}$ and ${\cal L}^{(0,1)}$ are the non-vacuum corrections to 
the particle energy and angular momentum given by the $\mathcal{O}(\epsilon)$ 
terms in Eqs.~\eqref{eq:circenergy}-\eqref{eq:circang}.

\end{appendix}

\bibliography{references.bib}

\end{document}